\documentclass[reprint,
superscriptaddress,
nofootinbib,
amsmath,
amssymb,
aps,
prd,
floatfix,
showkeys,
]{revtex4-2}

\usepackage{natbib}
\usepackage{graphicx,amsfonts,amssymb,amsbsy}
\usepackage{amsmath,amsthm,latexsym}
\usepackage{mathrsfs}
\usepackage[utf8]{inputenc} 
\usepackage{tabularx} 
\usepackage{multirow} 

\usepackage[normalem]{ulem}
\usepackage{fancyvrb}
\usepackage{fancyhdr}
\usepackage{placeins}
\usepackage{xcolor}
\usepackage[colorlinks]{hyperref}
\usepackage{calrsfs}    
\usepackage[scr=boondox]{mathalfa}
\usepackage[T1]{fontenc}
\usepackage{frcursive}

\usepackage{comment}

\usepackage{calligra}
\DeclareMathAlphabet{\mathcalligra}{T1}{calligra}{m}{n}
\DeclareFontShape{T1}{calligra}{m}{n}{<->s*[2.0]callig15}{}

\DeclareMathAlphabet{\mathpzc}{OT1}{pzc}{m}{it}

\begin{document}

\hypersetup{linkcolor=blue,citecolor=blue}

\title{Self-bound quark stars with a first-order two-to-three flavor phase transition}

\author{G. Teruya}
\email{giulliano.teruya@aluno.ufabc.edu.br}
\affiliation{Universidade Federal do ABC, Centro de Ci\^encias Naturais e Humanas, Avenida dos Estados 5001- Bang\'u, CEP 09210-580, Santo Andr\'e, SP, Brazil.}

\author{G. Lugones}
\email{german.lugones@ufabc.edu.br}
\affiliation{Universidade Federal do ABC, Centro de Ci\^encias Naturais e Humanas, Avenida dos Estados 5001- Bang\'u, CEP 09210-580, Santo Andr\'e, SP, Brazil.}

\author{A. G. Grunfeld}
\email{ag.grunfeld@conicet.gov.ar}
\affiliation{CONICET, Godoy Cruz 2290, Ciudad Autónoma de Buenos Aires, Argentina} 
\affiliation{Instituto de Astronomía y Física del Espacio (IAFE, CONICET-UBA), Intendente Güiraldes 2160 - Ciudad Universitaria, Ciudad Autónoma de Buenos Aires, Argentina}

\begin{abstract}
We investigate self--bound quark stars in a flavor--dependent quark--mass density--dependent model with an excluded--volume correction. We chart the parameter space at zero pressure to identify self--bound regimes, including parametrizations in which self--bound two--flavor matter undergoes a \emph{genuine first--order} $ud \to uds$ transition at finite pressure. 
We construct cold, $\beta$--equilibrated stellar sequences and compute the corresponding global properties (mass--radius relation, tidal deformability, and moment of inertia). For a wide region of the model parameter space, we find that the onset of a $uds$ core occurs \emph{before} the maximum--mass configuration is reached, yielding self--bound hybrid stars that follow the typical strange--quark--star sequence morphology but develop a characteristic \emph{kink} at $p_c=p_{\rm tr}$ along the stellar curves.
The excluded--volume parameter $\kappa$ controls the stiffness of the equation of state and thus masses, radii, tidal deformabilities, and moments of inertia; intermediate repulsion typically reconciles $M_{\max}\!\gtrsim\!2\,M_\odot$ with current astrophysical constraints. We further identify two equation--of--state--insensitive trends---dimensionless moment of inertia versus compactness and gravitational versus baryonic compactness. These results provide model--guided priors and tools for discriminating between hadronic and self--bound equations of state with multimessenger data.
\end{abstract}

\maketitle

\section{Introduction}

The standard description of neutron stars posits that they are composed of hadronic matter. Nevertheless, at sufficiently high baryon densities—such as those realized in neutron-star cores—strong-interaction dynamics may drive deconfinement and the emergence of quark degrees of freedom. Configurations hosting a quark-matter core are referred to as hybrid stars, whereas compact objects composed entirely of deconfined quark matter are known as strange quark stars when an absolutely stable, self-bound phase is realized.

Strange quark stars consist of strange quark matter (SQM), a self-bound deconfined phase of up, down, and strange quarks. The concept traces back to the seminal works of Itoh~\cite{Itoh:1970uw}, Bodmer~\cite{Bodmer:1971we}, and Witten~\cite{Witten:1984rs}, who proposed that the inclusion of the strange flavor can lower the energy per baryon below that of ordinary nuclear matter—the strange-matter hypothesis. In this picture, quark matter may constitute a stable phase with finite density at vanishing pressure. These considerations motivate a broader view of possible compact-star compositions and, in particular, prompt the question of whether self-bound phases might arise even in the \emph{absence} of strangeness.

Although the inclusion of the strange degree of freedom often lowers the energy per baryon—favoring a three-flavor self-bound phase—the viability of self-bound two-flavor phases cannot be dismissed \emph{a priori}. This has led to studies assessing whether two-flavor quark matter (udQM)—composed of up and down quarks only—could be absolutely stable and thus self-bound under specific conditions~\cite{Holdom:2017gdc}.  Even if \emph{bulk} udQM satisfies $E/A<930\,\mathrm{MeV}$ at $T=P=0$, this is not, in itself, incompatible with the stability of ordinary nuclei: finite-size contributions from surface and curvature energies penalize small quark droplets. Within the multiple–reflection–expansion (MRE) framework, these finite-size corrections raise the energy per baryon by amounts that scale as $\propto \sigma A^{-1/3}$ and $\propto \gamma A^{-2/3}$, respectively, where $A$ is the baryon number and $\sigma$ ($\gamma$) denotes the surface (curvature) tension~\cite{Madsen:1998uh}. As a result, nuclei can remain absolutely stable even if bulk udQM is the ground state: the decay of a nucleus into a finite $ud$-quark drop must overcome the surface and curvature barriers. Quantitatively, these effects imply a minimum baryon number $A_{\min}$ above which $ud$-quark matter droplets are favored, while hadronic matter is preferred below it; representative estimates yield $A_{\min}\!\gtrsim\!300$~\cite{Holdom:2017gdc}. The precise threshold depends on the microphysics that controls $\sigma$ and $\gamma$, for which effective-model calculations typically find sizable values, reinforcing the stabilizing role of finite-size terms~\cite{Lugones:2013ema,Lugones:2020qll}. These considerations parallel the classical strange-matter arguments~\cite{Witten:1984rs,Farhi:1984qu} while emphasizing that stability in bulk does not automatically extend to finite lumps. In the stellar context, macroscopic self-bound configurations can circumvent finite-size penalties and may be realized as compact stars composed of deconfined quark matter, whereas in vacuum small $ud$ droplets are unstable and hadronize.

Ab initio QCD at finite baryon density is presently out of reach: lattice calculations suffer from the sign problem, and perturbation theory is trustworthy only at asymptotically large densities. We therefore employ phenomenological equations of state (EOS). The MIT bag model encodes confinement through a bag constant $B$ but lacks dynamical chiral symmetry breaking/restoration, whereas the Nambu--Jona-Lasinio (NJL) model captures chiral dynamics via local four-fermion interactions yet does not confine~\cite{MITmodificado1,MITmodificado2,LugonesInterativo,NJL1,NJL2,Buballa:1998pr}. This complementarity motivates a framework that incorporates, in a minimal and consistent way, both confinement and (effective) chiral behavior. In this work we adopt the quark-mass density–dependent (QMDD) model, wherein quark effective masses acquire a density dependence. Enforcing thermodynamic consistency introduces additional terms in the pressure that act as a confining contribution, while the masses approach their current values at high density, mimicking chiral restoration~\cite{Lugones:2022upj,Lugones:2023zfd, Lugones:2024ryz}.

Two variants of the QMDD scheme were introduced in Ref.~\cite{Lugones:2022upj}: a \emph{flavor-blind} ansatz, where the effective mass depends on the baryon number density, and a \emph{flavor-dependent} ansatz, where each flavor’s mass tracks its own particle number density. A comprehensive survey of the parameter space shows that, in the flavor-blind case, among parameter sets that yield absolute stability, the vanishing-pressure ground state is invariably \emph{three}-flavor. In contrast, the flavor-dependent formulation admits a broad domain of parametrizations in which \emph{two}-flavor matter is self-bound at zero pressure \cite{Lugones:2022upj}.

{In the flavor-dependent scheme, parameter sets whose zero-pressure ground state is two-flavor \emph{necessarily} become three-flavor at sufficiently high pressure, since opening a new Fermi sea eventually renders the strange sector energetically favored. This provides a natural motivation to analyze how the $ud\!\to\!uds$ conversion proceeds and what sets its characteristic scales. In the flavor-dependent scheme, the dynamical masses can vary more independently than in the flavor-blind case, because each flavor responds to its own number density rather than to a common baryon-density scale. As a result, strangeness can be strongly suppressed at low densities while $u$ and $d$ quarks are not suppressed to the same extent, allowing self-bound two-flavor matter at $p\!\to\!0$ and delaying the onset of $s$ quarks to higher pressure. In fact, the exploratory study of Ref.~\cite{Lugones:2023zfd} indicates that the appearance of $s$ quarks need not be gradual: it can occur abruptly through a genuine first-order phase transition. This motivates a more systematic investigation of the $ud\!\to\!uds$ scenario and its astrophysical implications.}

Most of our previous astrophysical applications employed the flavor-blind variant~\cite{Lugones:2023zfd,Lugones:2024ryz,Lugones2025universality}. Here we focus on the flavor-dependent scheme and analyze, across the relevant stability window, the stellar sequences implied by parametrizations whose zero-pressure ground state is two-flavor. We solve for cold, $\beta$-equilibrated configurations and quantify how the ensuing quark--quark transition impacts global observables, in particular the mass--radius relation, tidal deformability, and moment of inertia, for the resulting \emph{self-bound hybrid stars}.
{Unlike conventional hybrid stars based on a hadron--to--quark phase transition, the stars considered here are self--bound at all pressures and contain only quark degrees of freedom. At the stellar level, the phase transition gives rise to configurations consisting of a $uds$ core surrounded by a $ud$ mantle.
This structure is qualitatively different from that obtained in models with a smooth crossover between quark phases, where no density discontinuity is present. The resulting self--bound hybrid stars occupy a distinct region of the mass--radius plane and exhibit characteristic behavior in their tidal deformabilities and related observables, which we analyze in detail below.}

{Very recently, hybrid quark stars featuring quark--quark phase transitions have also been investigated in Refs.~\cite{Yang:2025iyv, Zhang:2025rnf}.
In particular, Yang \emph{et al.} \cite{Yang:2025iyv} considered compact stars containing a $ud\to uds$ interface, while Zhang \emph{et al.} \cite{Zhang:2025rnf} explored related slow--stable (in the sense of Ref.~\cite{Lugones:2021bkm}) self--bound hybrid configurations within different effective descriptions.
Although these studies address a broadly similar physical scenario, their modeling strategy differs substantially from ours. Both works \cite{Yang:2025iyv, Zhang:2025rnf} rely on highly parameterized, piecewise prescriptions for the EOS, in which the transition is implemented  between two independently modeled phases. {As a result, key characteristics—such as the transition pressure and the associated energy--density discontinuity—are not derived from a single, unified microphysical framework, and the origin of the $ud\to uds$ threshold is not traced back to an underlying dynamical mechanism.}
By contrast, in our framework the $ud \to uds$ transition emerges dynamically from the particle--number--density dependence of the quark masses, without introducing a separate phase--transition construction by hand. Moreover, we quantify how excluded--volume effects reshape the stellar sequences and identify equation--of--state--insensitive universal relations specific to self--bound hybrid stars. These aspects allow us to complement and extend the recent literature by
highlighting both the microphysical origin of the transition and its distinct astrophysical implications.}

The paper is organized as follows. Section \ref{cap2} introduces the flavor-dependent QMDD framework, detailing the density-dependent mass ansatz and the excluded-volume prescription, and summarizes the resulting zero-temperature EOS. In Section \ref{sec:stability_window} we delineate the stability window in the $(a,C)$ plane and discuss causality bounds. Section \ref{cap4} presents the thermodynamics and phase structure—Gibbs energy per baryon, $p(\epsilon)$, and speed of sound—quantifying the two- to three-flavor transition (transition pressure and energy-density jump). 
In Section~\ref{cap5} we construct cold, $\beta$-equilibrated stellar sequences by integrating the Tolman--Oppenheimer--Volkoff equations and the standard perturbation equations for tidal deformability and slow rotation (moment of inertia), and we compare the resulting mass--radius relations with current multimessenger constraints.  Section \ref{sec:universal_relations} discusses EOS-insensitive (“universal”) relations for self-bound stars. 
We conclude by summarizing the main results and their implications in Sec. \ref{sec:conclusions}. Technical details of the EOS construction and of the computations of Love numbers and the moment of inertia are collected in the Appendices \ref{sec:appendix_EOS} and \ref{sec:appendix_stellar_structure}.

\section{The flavor-dependent QMDD model}
\label{cap2}

We model cold quark matter as a gas of non-interacting quasiparticles of flavors $u,d,s$ with effective masses $M_i$, supplemented by free electrons to ensure charge neutrality. Within the QMDD framework~\cite{Lugones:2022upj}, interaction effects are encoded through density-dependent quark masses while the expressions for the energy density and the number densities retain the free-gas functional form. Specifically, the effective mass of each flavor is taken as
\begin{equation}
M_i = m_i + \frac{C}{n_i^{a/3}},
\label{eq:formula_1}
\end{equation}
where $i\in\{u,d,s\}$, $m_i$ is the (constant) current mass of flavor $i$, $n_i$ is its number density, and $C$ and $a$ are considered flavor-independent parameters.

{We note that the flavor-dependent QMDD scheme already embodies $SU(3)_f$ breaking at the dynamical level through the strange-quark current mass and the resulting distinct density evolution of the $s$ sector under $\beta$ equilibrium and charge neutrality. In this sense, the model is “flavor-dependent” because each quasiparticle mass tracks the number density of its own flavor, $M_i=M_i(n_i)$, rather than a common baryon-density scale as in flavor-blind variants. This reduced correlation among the density dependences of $M_u$, $M_d$, and $M_s$ plays a central role in enabling domains where strangeness is suppressed at low pressure while $u$ and $d$ quarks are not, thereby opening the possibility that $ud$ matter is self-bound at $p\!\to\!0$ and that the onset of $s$ quarks is delayed to higher pressures.}

{At the same time, Eq.~\eqref{eq:formula_1} adopts a minimal parametrization in which the mass-functional parameters $(a,C)$ are taken to be flavor-independent. Allowing explicit parameter flavor dependence, e.g., $C_s\neq C_{u,d}$ and/or $a_s\neq a_{u,d}$, would further enlarge the model space and may shift quantitative outcomes such as the transition density/pressure and the size of the $uds$ core. We do not pursue that extension here in order to keep the parameter survey tractable and to isolate the physical consequences of the flavor-dependent density evolution itself; explicit flavor dependence in $(a,C)$ constitutes a natural next step beyond the scope of the present work (see also the exploratory study in the Appendix of Ref.~\cite{Lugones:2022upj}).
}

Units are chosen so that $M_i$ is in MeV when $n_i$ is expressed in $\mathrm{fm}^{-3}$; throughout we adopt natural units with $\hbar=c=1$. In contrast to our earlier “flavor-blind” formulation—where $M_i$ depended only on the baryon number density—we employ here a flavor-dependent mass function $M_i(n_i)$.

The limiting behavior of Eq.~\eqref{eq:formula_1} reproduces the expected qualitative features: as $n_i\!\to\!0$, $M_i$ diverges, suppressing quasiparticle excitations and thereby mimicking confinement; at asymptotically large densities, $M_i\!\to\!m_i$, providing a phenomenological realization of (effective) chiral restoration, particularly if $m_i=0$. The exponent $a$ can be related to a schematic interquark potential, $v(r)\propto r^{a}$, which controls how rapidly medium effects diminish with increasing density~\cite{Lugones:2022upj}.

A natural way to encode repulsive interactions in the flavor-dependent QMDD model is to introduce a phenomenological excluded-volume (EV) correction that penalizes arbitrarily high local number densities, thereby stiffening the EOS at moderate and high densities~\cite{Lugones:2023zfd}. {In the absence of an underlying Lagrangian with explicit mediator fields, the EV prescription provides an economical effective parametrization of short-range repulsive channels that are expected to be operative in dense, nonperturbative QCD matter. Such repulsion is also indicated empirically by the existence of compact stars with masses $\gtrsim 2\,M_\odot$, which generically requires a sufficiently stiff high-density EOS. In this sense, EV plays here a role analogous to vector interactions in field-theoretical descriptions (e.g., relativistic mean-field models for hadronic matter or NJL/vector-MIT extensions for quark matter), while remaining consistent with the defining structure of QMDD models based on density-dependent mass functionals.}

Consistent with the flavor-dependent ansatz in Eq.~\eqref{eq:formula_1}, we take quarks of flavor $i$ to reduce only the volume available to the same flavor (no interspecies reduction):
\begin{equation}
\tilde{V}_i  =  V - b(n_i)\,n_i\,V  =  V\bigl[1 - b(n_i)\,n_i\bigr],
\label{eq:formula_2}
\end{equation}
where $b(n_i)$ is the effective excluded volume per particle of flavor $i$ (dimension of volume).

{We set $b(n_i)$ via an ansatz that captures the expected qualitative behavior of an effective repulsion: it suppresses excessively large local densities and increases the pressure at a given energy density, thereby stiffening the EOS at moderate densities, while becoming negligible at very high densities, in line with asymptotic freedom. In the present implementation, the repulsive correction is incorporated through the available-volume factor in Eq.~\eqref{eq:formula_2}, i.e., through a finite-volume (``hard-core'') effect akin to Van der Waals–type treatments.}
{For consistency with the flavor-dependent mass prescription in Eq.~\eqref{eq:formula_1}, we also allow the excluded volume per particle to depend on the density of the same flavor.}
Accordingly, in analogy with Refs.~\cite{Lugones:2023zfd,Lugones:2022upj}, we adopt
\begin{equation}
b_i = \frac{\kappa}{n_i},
\end{equation}
with $\kappa>0$. Substituting this ansatz into Eq.~\eqref{eq:formula_2} gives $\tilde{V}_i = V\,(1 - \kappa)$ for $i = u, d, s$. Thus, in spite of the flavor-dependent parametrization, this ansatz leads to the same repulsive contribution across flavors, making the index $i$ superfluous:
\begin{eqnarray}
\tilde{V}_i  \equiv  \tilde{V} =  V\,(1 - \kappa).
\label{eq:formula_3}
\end{eqnarray}
 
{Substituting Eqs.~\eqref{eq:formula_1} and \eqref{eq:formula_3} into the general flavor-dependent QMDD expressions of Ref.~\cite{Lugones:2023zfd} yields the complete EOS, in which the EV correction effectively mimics repulsive interactions without introducing additional mediator fields and couplings.}
We collect the resulting formulas in Appendix~\ref{sec:appendix_EOS}.

\section{Stability window}
\label{sec:stability_window}

At zero pressure, the energy per baryon
\begin{equation}
e^0  \equiv  \left.\frac{\epsilon}{n_B}\right|_{p=0},
\end{equation}
diagnoses whether quark matter is self-bound or only realized at high pressure. We take $930\,\mathrm{MeV}$—the energy per nucleon of the most tightly bound nucleus, $^{62}\mathrm{Ni}$—as the reference benchmark. If $e^0<930\,\mathrm{MeV}$, bulk hadronization is energetically disfavored and the matter is self-bound, so compact stars composed of it are pure quark stars. If $e^0>930\,\mathrm{MeV}$, quark matter is unbound at low pressure and the EOS is hybrid: hadronic at low pressure and deconfined only beyond a transition pressure, yielding hybrid stars with a quark core and a hadronic envelope.

\begin{figure}[tb]
\includegraphics[width=1\columnwidth]{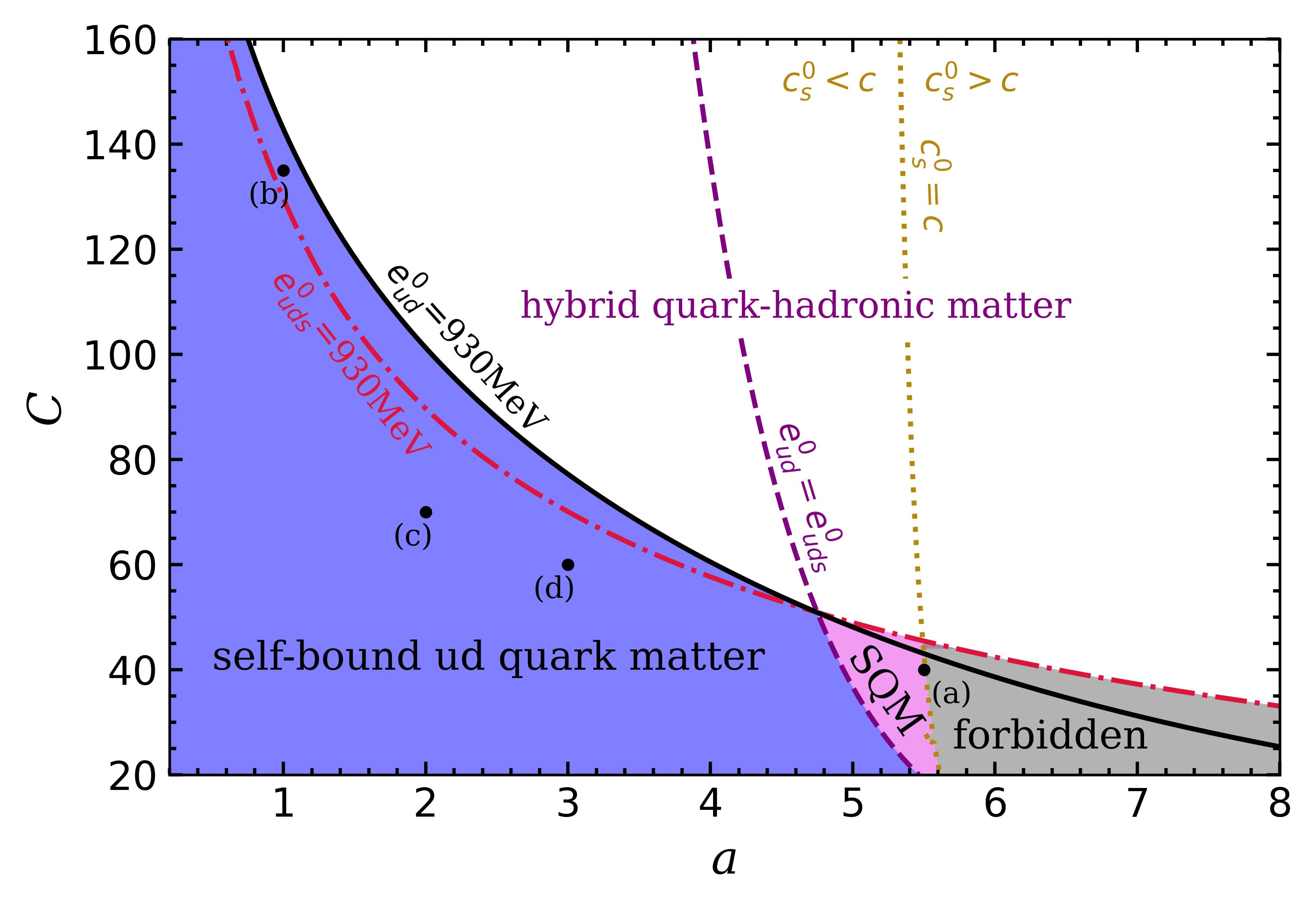}
\caption{Stability window of the flavor-dependent QMDD model in the $(a,C)$ plane. The black (red) curve corresponds to $e_{ud}^0=930\,\mathrm{MeV}$ ($e_{uds}^0=930\,\mathrm{MeV}$); points lying below the respective curve yield self-bound $ud$ ($uds$) matter at $p=0$. Points above both 930\,MeV curves correspond to hybrid quark-hadronic matter. The indigo line marks $e_{ud}^0=e_{uds}^0$. The orange line indicates $c_s^0=c$; to its right the $p\!\to\!0$ limit is acausal.}
\label{fig:1}
\end{figure}

Figure~\ref{fig:1} shows the stability window of the flavor-dependent QMDD model in the $(a,C)$ plane. The curves $e_{ud}^0=930\,\mathrm{MeV}$ (black) and $e_{uds}^0=930\,\mathrm{MeV}$ (red) delimit where two- and three-flavor matter, respectively, are self-bound at zero pressure: points lying below the former (latter) yield self-bound $ud$ ($uds$) matter. Accordingly, self-bound $ud$ quark matter (udQM) is realized when $e_{ud}^0<e_{uds}^0$, whereas SQM occurs for parameter choices with $e_{uds}^0<e_{ud}^0$. Points above both 930\,MeV curves correspond to hybrid EOSs—hadronic at low pressure and deconfined only beyond a transition pressure—leading to stars with a quark core and a hadronic envelope. An additional curve, $c_s^0=c$, marks parameter choices for which the sound speed at zero pressure equals the speed of light. To its left, the EOS remains causal down to $p=0$; to its right, the $p\!\to\!0$ limit becomes acausal ($c_s^0>c$), thereby excluding self-bound quark matter.
{Because excluded-volume repulsion stiffens the EOS at finite density, one might expect it to also affect the $c_s^0=c$ boundary. In our implementation, however, the $p\!\to\!0$ sound speed is \emph{independent} of the excluded-volume  (cf. Ref.~\cite{Lugones:2023zfd}): we have verified numerically that $c_s^0$ is invariant under changes in $\kappa$, so the causality line in Fig.~\ref{fig:1} is determined solely by $(a,C)$.}
{This behavior is consistent with the structure of the EV prescription: the quark-sector thermodynamics is obtained from the pointlike expressions by a constant available-volume factor and a density rescaling, which leaves the ratio $dp/d\epsilon$ unchanged in the $p\!\to\!0$ limit. Consequently, varying $\kappa$ does not expand the acausal region and does not alter whether a given benchmark point is excluded by the $c_s^0>c$ criterion.}
Therefore, although the flavor-dependent QMDD model involves three parameters—$a$, $C$, and the excluded-volume parameter $b$, the stability region is  fully determined by $(a,C)$.

The purple region in Fig.~\ref{fig:1} corresponds to parameter choices for which bulk two-flavor quark matter would be more bound than the most tightly bound nucleus. As noted in the Introduction, this does not necessarily entail an instability of ordinary nuclei: finite-size costs from surface and curvature terms disfavor the formation of small $ud$ droplets, so nuclear stability is preserved. The pink region (SQM) is likewise compatible with nuclear stability.

We analyze four benchmark points in the $(a,C)$ plane, indicated by (a)–(d) in Fig.~\ref{fig:1} and listed in Table~\ref{tab:param_sets}. Point (a), $(a,C)=(5.5,40)$, lies in the SQM sector and serves as the reference strange–quark–matter EOS. Points (b) $(1,135)$, (c) $(2,70)$, and (d) $(3,60)$ sample the self-bound $ud$ domain and exemplify self-bound hybrid configurations in which a two– to three–flavor transition can occur. Unless stated otherwise, each benchmark is evaluated at several excluded–volume strengths, typically $\kappa=0,\,0.3,\,0.6$~\cite{Lugones:2023zfd}; the chosen $\kappa$ values will be specified in the corresponding figures and captions.

\begin{table}[tb]
\caption{Benchmark parameter sets used in this work. Labels (a)–(d) correspond to the points highlighted in Fig.~\ref{fig:1}. Model (a) is a pure strange–quark–matter case; models (b)–(d) are self–bound hybrid configurations featuring a two– to three–flavor ($ud\!\to\!uds$) transition.}
\label{tab:param_sets}
\begin{ruledtabular}
\begin{tabular}{ccc}
Label in Fig.~\ref{fig:1} & $a$ & $C$ \\
\colrule
(a) & 5.5 & 40  \\
(b) & 1   & 135 \\
(c) & 2   & 70  \\
(d) & 3   & 60  \\
\end{tabular}
\end{ruledtabular}
\end{table}

\section{NUMERICAL RESULTS FOR THE EOS} 
\label{cap4}

{Before presenting the detailed numerical results, it is useful to clarify the physical origin of the $ud\!\to\!uds$ transition in the present framework. In the flavor-dependent QMDD formulation, each dynamical quark mass tracks the number density of its own flavor, $M_i=M_i(n_i)$, rather than a common baryon-density scale as in flavor-blind variants. This weakens the correlation among the density evolutions of $M_u$, $M_d$, and $M_s$, so that strangeness can remain strongly suppressed at low densities (large $M_s$) while $u$ and $d$ quarks are not suppressed to the same extent. Combined with our excluded-volume prescription, this mechanism naturally generates a stability window in which self-bound two-flavor matter exists at $p\!\to\!0$ even though three-flavor matter is not yet favored. As the density increases, $M_s$ decreases and $uds$ matter eventually becomes thermodynamically preferred; the conversion occurs when the Gibbs free energies cross, producing a genuine first-order transition with a finite density (energy-density) jump. Importantly, the transition is not imposed by hand, but emerges dynamically from the density dependence of the quark masses.}

\begin{figure}[tb]
\centering
\includegraphics[width=0.99\linewidth]{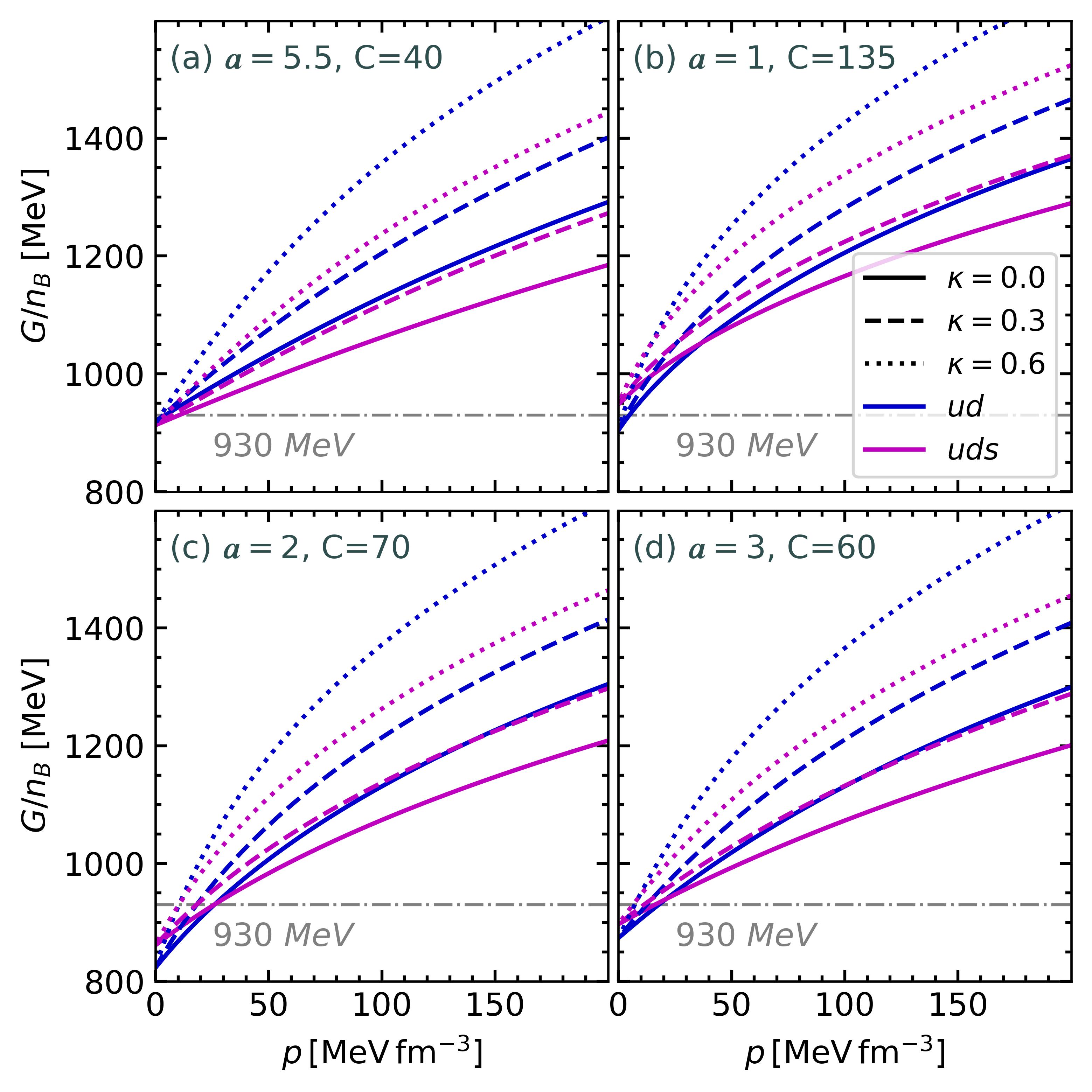}
\caption{Gibbs free energy per baryon for two- ($ud$) and three-flavor ($uds$) bulk quark matter. Curves correspond to the parameter sets (a)–(d) defined in Fig.~\ref{fig:1} and three representative excluded-volume strengths $\kappa$ (0, intermediate, high). At $p=0$, $G/n_B$ is independent of $\kappa$. Crossings between $ud$ and $uds$ branches (when present) indicate a first-order $ud\!\to\!uds$ transition.}
\label{fig:2}
\end{figure}

Figure~\ref{fig:2} displays the Gibbs free energy per baryon, $G/n_B=(\epsilon+p)/n_B$, for the parameter sets (a)–(d) defined in Fig.~\ref{fig:1}. For each $(a,C)$, we show three representative excluded-volume strengths $\kappa$ (0, intermediate, high) as specified in Fig.~\ref{fig:1}. At $p=0$, $G/n_B$ is independent of $\kappa$ for both $ud$ and $uds$ matter. In all four cases, at least one of the two compositions satisfies $G/n_B<930\,\mathrm{MeV}$ at zero pressure.

Two different scenarios emerge from these results. In Fig.~\ref{fig:2}(a), the $uds$ branch lies below the corresponding $ud$ branch at all pressures, identifying the strange-quark-matter scenario in which three-flavor matter is favored throughout. By contrast, Figs.~\ref{fig:2}(b)–\ref{fig:2}(d) exhibit a scenario in which $ud$ matter minimizes $G/n_B$ at low pressure, whereas $uds$ becomes energetically favored above a threshold. The ensuing $ud\!\to\!uds$ conversion is first order, as indicated by the crossing of the $G/n_B$ branches and the accompanying discontinuity in energy density. The transition pressure depends on $(a,C)$ and decreases monotonically with increasing $\kappa$; the three panels illustrate representative cases in which the transition occurs at low, moderate, and high pressures. As discussed below, this systematic dependence on $\kappa$ directly influences the extent of the three-flavor core in self-bound hybrid stars under different excluded-volume choices.

\begin{figure}[tbh]
\centering
\includegraphics[width=0.99\linewidth]{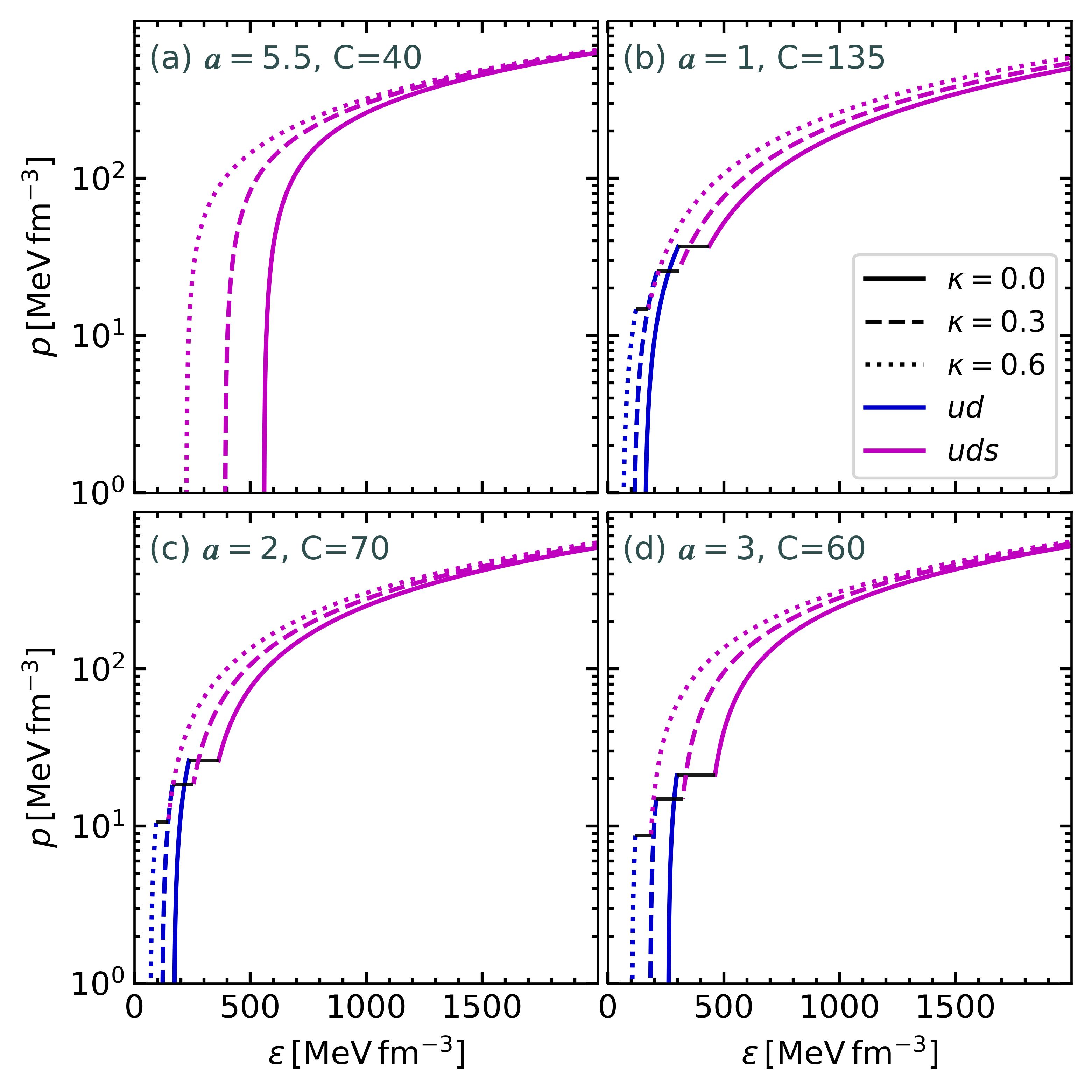}
\caption{Total pressure $p$ versus energy density $\epsilon$ for the same parameter sets as Fig.~\ref{fig:2}. A constant-pressure plateau with a discontinuous jump in $\epsilon$ appears when a first-order $ud\!\to\!uds$ transition occurs; each curve crosses $p=0$ at finite $\epsilon$.}
\label{fig:3}
\end{figure}

\begin{table*}[tb]
\caption{{First-order $ud\!\to\!uds$ transition parameters for the flavor-dependent QMDD EOS. 
For each $(a,C,\kappa)$ we list the transition pressure $p_\mathrm{tr}$, the energy densities just below/above the transition ($\epsilon_{-}$, $\epsilon_{+}$), and the jump $\Delta\epsilon\equiv\epsilon_{+}-\epsilon_{-}$. 
The last column reports the Seidov onset-instability diagnostic
$\gamma$ defined in Eq.~\eqref{eq:diagnostic_Seidov}; values $\gamma<1$ indicate that Seidov's threshold is not reached at the transition onset. 
We emphasize that the definitive stability assessment is obtained from the TOV equilibrium sequences (Sec.~\ref{cap5}), and we use $\gamma$ only as an auxiliary check.}  }
\label{table_2}
\begin{ruledtabular}
\begin{tabular}{cccccccc}
$a$ & $C$ & $\kappa$
& $p_\mathrm{tr}$ (MeV\,fm$^{-3}$)
& $\epsilon_{+}$ (MeV\,fm$^{-3}$)
& $\epsilon_{-}$ (MeV\,fm$^{-3}$)
& $\Delta\epsilon$ (MeV\,fm$^{-3}$)
& $\gamma$ \\
\colrule
1 & 135 & 0.0 & 36.8 & 436.7 & 303.8 & 132.9 & 0.855 \\
1 & 135 & 0.3 & 25.8 & 305.8 & 212.7 & 93.0 & 0.855 \\
1 & 135 & 0.6 & 14.7 & 174.8 & 121.7 & 53.1 & 0.854 \\
\colrule
2 & 70 & 0.0 & 26.2 & 363.4 & 235.8 & 127.6 & 0.925 \\
2 & 70 & 0.3 & 18.4 & 254.5 & 165.2 & 89.3 & 0.924 \\
2 & 70 & 0.6 & 10.6 & 145.6 & 94.5 & 51.1 & 0.924 \\
\colrule
3 & 60 & 0.0 & 21.1 & 462.5 & 297.5 & 165.0 & 0.968 \\
3 & 60 & 0.3 & 14.9 & 324.0 & 208.3 & 115.7 & 0.968 \\
3 & 60 & 0.6 & 8.7 & 185.6 & 119.2 & 66.4 & 0.967 \\
\end{tabular}
\end{ruledtabular}
\end{table*}

Figure~\ref{fig:3} shows the total pressure $p$ as a function of the energy density $\epsilon$ for the same parameter sets as in Fig.~\ref{fig:2}. In all cases, $p(\epsilon)$ crosses zero at a finite $\epsilon$, reflecting the effective bag-like vacuum contribution generated by the density dependence of the quasiparticle masses in the QMDD model. For EOSs that undergo a first-order $ud\!\to\!uds$ transition, $p(\epsilon)$ exhibits a constant-pressure plateau accompanied by a discontinuous jump in $\epsilon$ (latent heat).

Table~\ref{table_2} summarizes the transition pressure $p_\mathrm{tr}$ and the associated energy densities. For fixed $(a,C)$, increasing $\kappa$ lowers $p_\mathrm{tr}$ and the energy-density discontinuity $\Delta\epsilon$; this monotonic dependence will be relevant for the stellar sequences discussed below.

{Seidov~\cite{Seidov1971} derived a useful \emph{onset-instability} diagnostic for first-order phase transitions in general relativity: a spherically symmetric configuration becomes unstable \emph{at the transition onset} if the energy-density discontinuity across the interface exceeds the critical threshold
\begin{equation}
\frac{\epsilon_{+}}{\epsilon_{-}} =  \frac{3}{2}\!\left(1+\frac{p_\mathrm{tr}}{\epsilon_{-}}\right),
\label{eq:seidov_parameter}
\end{equation}
where $p_\mathrm{tr}$ is the transition pressure and $\epsilon_{-}$ ($\epsilon_{+}$) is the energy density just below (above) the transition.
For convenience, Table~\ref{table_2} lists $p_\mathrm{tr}$, $\epsilon_{-}$, and $\epsilon_{+}$ for the benchmark EOSs considered here and reports in the last column the corresponding diagnostic ratio
\begin{equation}
\gamma \equiv \frac{\epsilon_{+}/\epsilon_{-}} {\tfrac{3}{2}\bigl(1+p_\mathrm{tr}/\epsilon_{-}\bigr)}\, .
\label{eq:diagnostic_Seidov}
\end{equation}
We find $\gamma<1$ for all benchmark transitions shown (typically $0.85$--$0.97$), indicating that Seidov's onset-instability threshold is not reached in these cases.
We stress, however, that this criterion is only an auxiliary check: the definitive stability assessment is obtained by constructing the full TOV sequences and applying the turning-point condition. We confirm in Sec.~\ref{cap5} that the sequences containing a $ud\!\to\!uds$ interface exhibit dynamically stable hybrid branches.}

{Table~\ref{table_2} further shows that, for fixed $(a,C)$, the individual transition quantities $p_\mathrm{tr}$, $\epsilon_{-}$, and $\epsilon_{+}$ vary appreciably with $\kappa$, whereas the Seidov diagnostic $\gamma$ is essentially insensitive to $\kappa$ within each family.
Finally, we have verified that $\gamma$ remains below unity throughout the \emph{causal self-bound} region of the $(a,C)$ parameter space explored in Fig.~\ref{fig:1}; since this diagnostic is not central to the main goals of this work, we do not include an exhaustive scan here, but note it for completeness.}

\begin{figure}[tbh]
\centering
\includegraphics[width=0.99\linewidth]{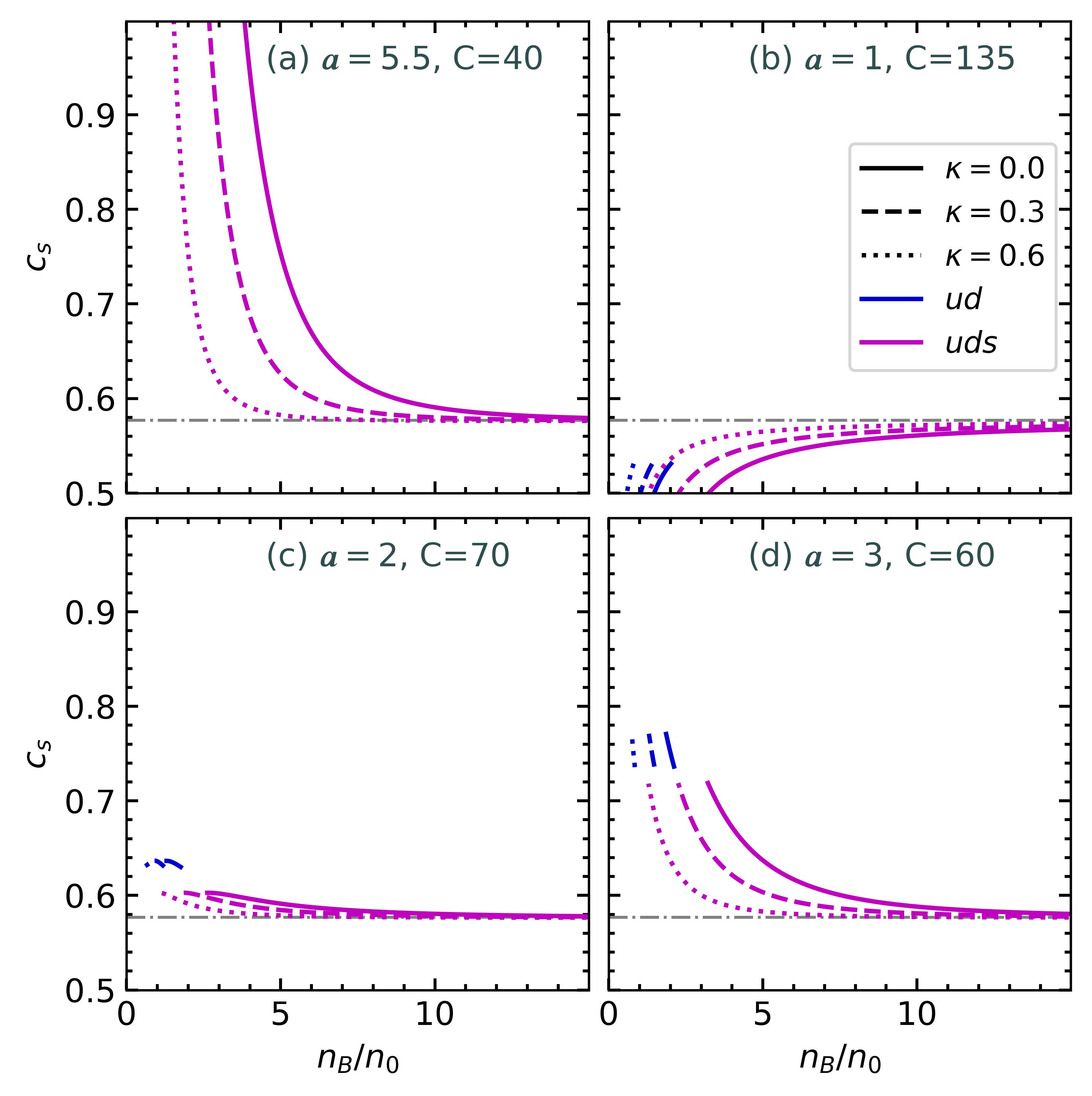}
\caption{Sound speed $c_s$ (in units of $c$) versus baryon number density $n_B$ for the same parameter sets as Fig.~\ref{fig:2}. The horizontal line marks the conformal limit $1/\sqrt{3}$; curves end at their zero-pressure points.}
\label{fig:4}
\end{figure}

Figure~\ref{fig:4} shows the sound speed $c_s$ (in units of $c$) for the same parameter sets as in Fig.~\ref{fig:2}. At large baryon density $n_B$, $c_s$ approaches the conformal limit $1/\sqrt{3}$. Each curve terminates at its zero-pressure point; the associated termination densities depend on $\kappa$. In Figs.~\ref{fig:4}(b)–\ref{fig:4}(d), the first-order $ud\!\to\!uds$ transition produces a constant-pressure interval with a discontinuous energy density, so $c_s$ is undefined there and the interval is omitted.

\section{Global properties of Hybrid self-bound quark stars}
\label{cap5}

\begin{figure}[tb]
\centering
\includegraphics[width=0.97\linewidth]{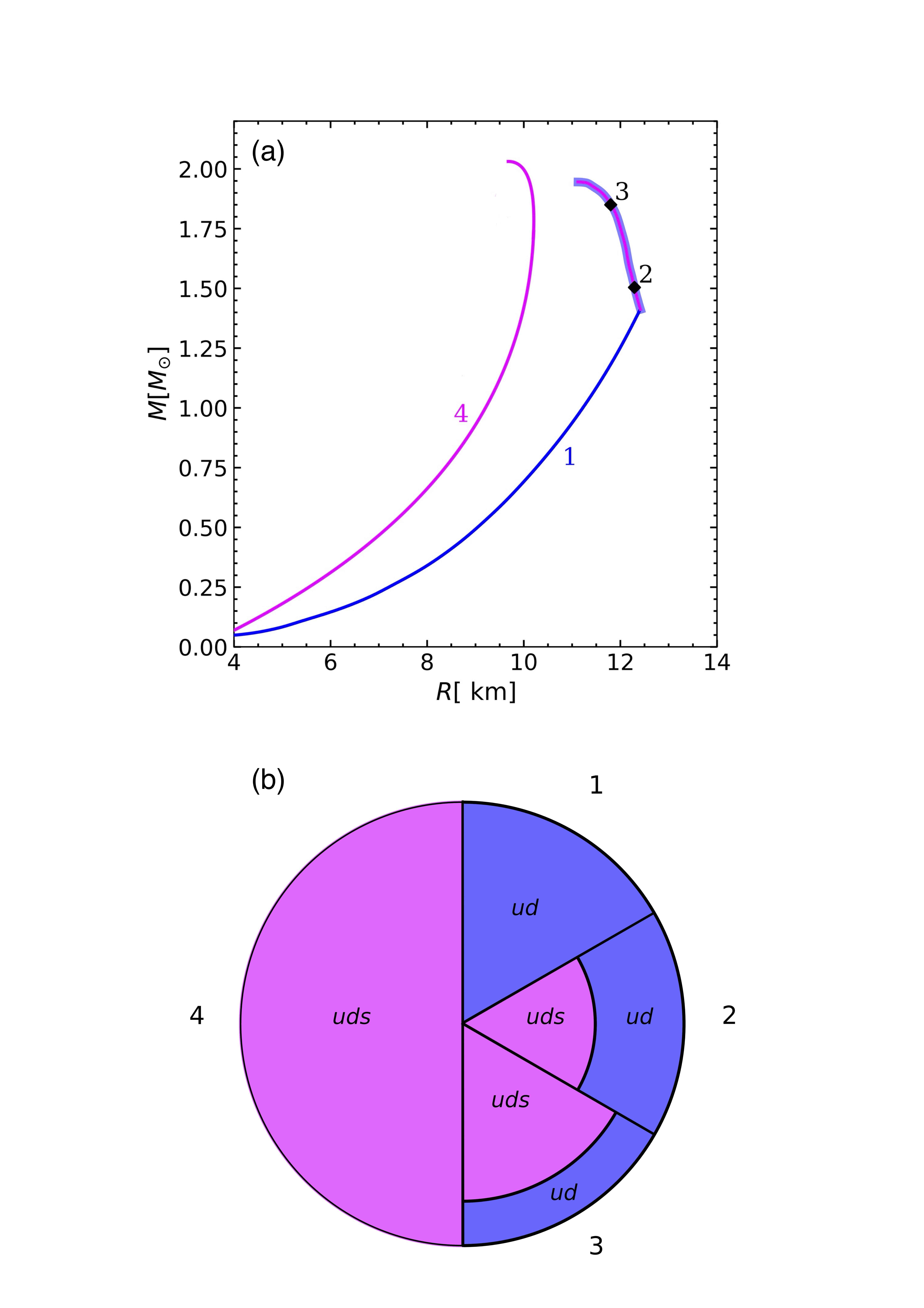}
\caption{(a) Schematic mass–radius diagram of self-bound stars. The \textit{magenta} curve shows self-bound strange-quark-matter configurations, the \textit{blue} curve shows self-bound $ud$ stars, and the \textit{bicolor} (magenta–blue) curve displays the sequence of self-bound hybrid stars analyzed in this work. (b) Representation of the internal composition of the models on the $M$–$R$ curves in panel (a), indicating regions of $uds$ matter (magenta) and $ud$ matter (blue).}
\label{fig:5}
\end{figure}

Having established the EOS properties across the chosen parameter sets, we now turn to the stellar structure they imply. Equilibrium configurations are obtained by integrating the general-relativistic TOV equations (see Appendix~\ref{sec:appendix_TOV}). As shown below, the global properties are highly sensitive to $(a,C,\kappa)$, giving rise to distinct classes of self-bound compact stars.
The parameter-space regions delineated in Fig.~\ref{fig:1} lead to qualitatively different sequences, summarized schematically in Fig.~\ref{fig:5}. Parameters in the SQM domain yield stars composed entirely of $uds$ matter (configuration “4”). Parameters in the self-bound $ud$ region ensure that the low-density end of the sequence (near $p\!\to\!0$) remains two-flavor, producing the configurations labeled “1”, “2”, and “3” in Fig.~\ref{fig:5} according to the central pressure $p_c$ (the pressure at the stellar center). Specifically, if $p_c<p_\mathrm{tr}$ the star is purely two-flavor (configuration “1”), whereas for $p_c>p_\mathrm{tr}$ a $uds$ core forms within a $ud$ mantle (configurations “2” and “3”), i.e., a self-bound hybrid star.
For clarity, we denote by $p_c^{\max}$ the central pressure of the configuration that attains the maximum gravitational mass $M_{\max}$ along a given TOV sequence. Numerically, we do not encounter sequences with $p_c^{\max}<p_\mathrm{tr}$. For every self-bound choice in Fig.~\ref{fig:1}, the onset of the $uds$ core precedes the maximum-mass configuration ($p_c^{\max}\!\ge\!p_\mathrm{tr}$), and the “purely two-flavor up to $M_{\max}$” scenario is excluded by our solutions.

\begin{figure}[tb]
\centering    
\includegraphics[width=0.99\linewidth]{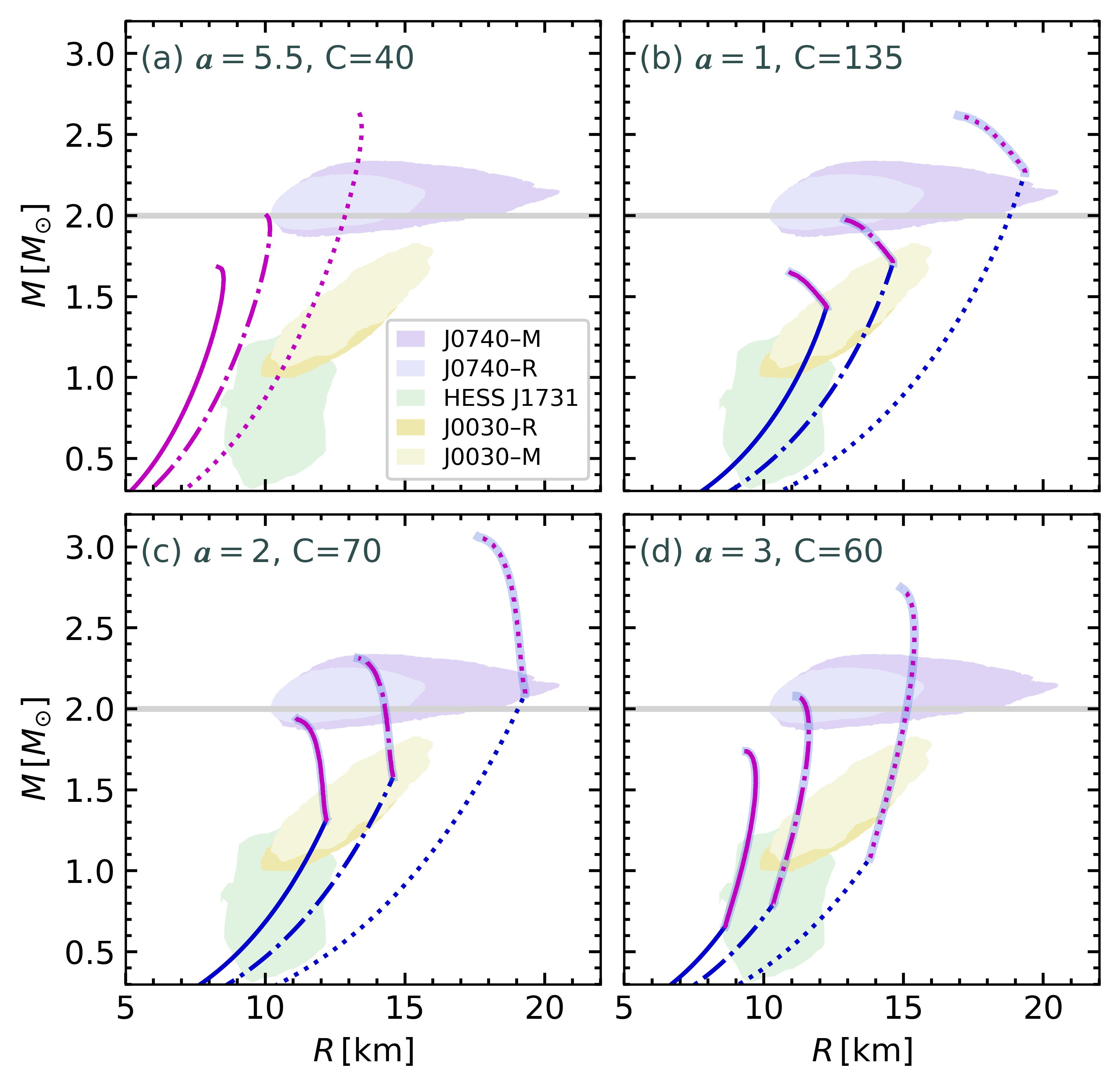}
\caption{Mass–radius relations for quark-star sequences computed with the EOS sets of Table~\ref{tab:param_sets}. Line styles encode the excluded-volume strength $\kappa$ (solid, dashed, dotted), following the convention of Fig.~\ref{fig:2}. Colors indicate composition: magenta—pure $uds$ (SQM); blue—pure $ud$; bicolor—self-bound hybrids with a $uds$ core surrounded by a $ud$ mantle. (a) Parameters in the SQM domain yield purely three-flavor sequences. (b)–(d) EOSs featuring a first-order $ud\!\to\!uds$ transition placed at low, intermediate, and high densities, respectively; the kink marks $p_c=p_\mathrm{tr}$. Increasing $\kappa$ stiffens the EOS and raises $M_{\max}$. Shaded contours denote observational constraints from  PSR~J0030+0451~\cite{Riley2019,Miller2019}, HESS~J1731$-$347~\cite{Doroshenko2022}, and PSR~J0740+6620~\cite{Riley2021,Miller2021}.}
\label{fig:6}
\end{figure}

Using the same EOS parameter sets as above, we integrate the TOV equations and compute the stellar mass $M$, radius $R$, tidal deformability $\Lambda$, and moment of inertia $I$.

Figure~\ref{fig:6} presents the mass–radius relations. Within each panel, line styles encode the excluded-volume strength $\kappa$ (solid, dashed, dotted), following the same convention adopted in Fig.~\ref{fig:2} (and subsequent figures). Increasing $\kappa$ stiffens the EOS and raises $M_{\max}$. Panel~\ref{fig:6}(a) employs parameters in the SQM domain of Fig.~\ref{fig:1}, yielding sequences of pure three-flavor strange-quark stars (magenta). Panels~\ref{fig:6}(b)–\ref{fig:6}(d) use EOSs with a first-order $ud\!\to\!uds$ transition; the three parameter sets place the transition at low, intermediate, and high densities, respectively. For central pressures below the transition ($p_c<p_\mathrm{tr}$), the sequence consists of two-flavor stars (blue). Once $p_c>p_\mathrm{tr}$, a $uds$ core forms within a $ud$ mantle, producing self-bound hybrid stars (bicolor magenta/blue). The kink at $p_c=p_\mathrm{tr}$ marks the onset of the core.

Figure~\ref{fig:6} confronts the mass–radius sequences with current astrophysical constraints. A clear dependence on the excluded-volume strength is evident. Models with $\kappa=0$ are too soft: their $M$–$R$ curves fail to reach $2\,M_\odot$, underscoring the need for repulsive interactions in the quark sector. At the other extreme, $\kappa=0.6$ produces very stiff EOSs. Although these sequences comfortably attain large maximum masses, they generally overpredict radii; in panels~\ref{fig:6}(b)–\ref{fig:6}(d) the curves lie on or above the upper edge of the PSR~J0030+0451 and PSR~J0740+6620 constraints. An exception occurs in panel~\ref{fig:6}(a), where the pure SQM models with $\kappa=0.6$ are compatible with all bands, a marked improvement over $\kappa=0$ and $0.3$. For the self-bound hybrid cases, the intermediate choice $\kappa=0.3$ provides the best overall agreement, balancing the $2\,M_\odot$ requirement against radius constraints. Across panels~\ref{fig:6}(b)–\ref{fig:6}(d), the predicted $M_{\max}$ varies with the transition density: it is lower for both low and high transition densities and peaks for the intermediate case in panel~\ref{fig:6}(c).

\begin{figure}[tb]
\centering
\includegraphics[width=0.99\linewidth]{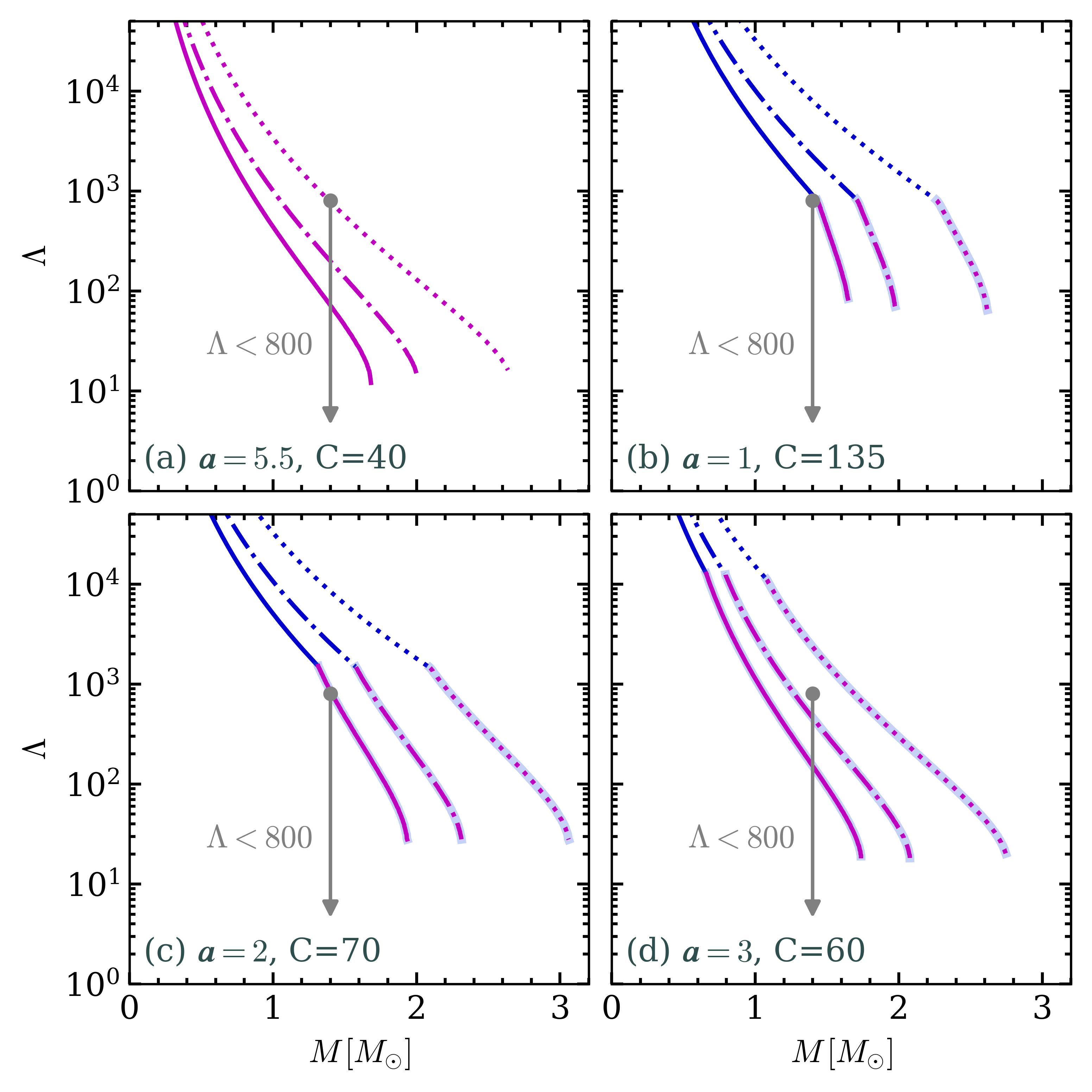}
\caption{Dimensionless tidal deformability $\Lambda$ as a function of gravitational mass $M$ for the same EOS sets used in Fig.~\ref{fig:6}. Line styles encode the excluded-volume strength $\kappa$ (solid, dashed, dotted), following the convention of Fig.~\ref{fig:2}. (a) Pure strange–quark–star sequences; (b)–(d) self-bound hybrid-star sequences with the $ud\!\to\!uds$ transition placed at low, intermediate, and high densities, respectively. }
\label{fig:7}
\end{figure}

\begin{figure}[tb]
\centering
\includegraphics[width=0.99\linewidth]{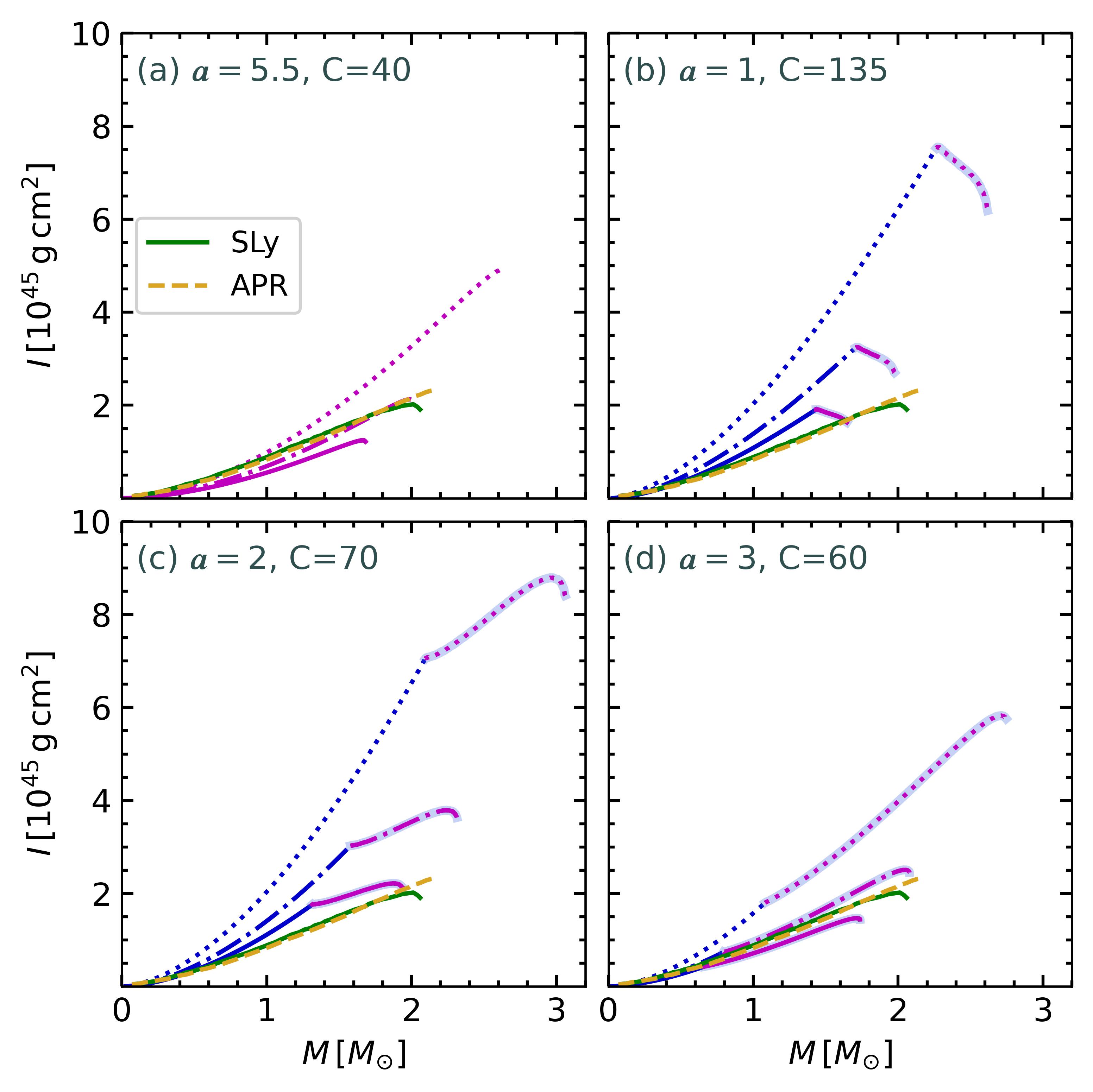}
\caption{Moment of inertia $I$ as a function of gravitational mass $M$ for the same EOS sets used in Fig.~\ref{fig:6}.  (a) Pure strange–quark–star sequences; (b)–(d) self-bound hybrid-star sequences with the $ud\!\to\!uds$ transition placed at low, intermediate, and high densities, respectively. For reference, hadronic benchmarks (SLy and APR) are overplotted. }
\label{fig:8}
\end{figure}

{
Figure~\ref{fig:7} displays the (dimensionless) tidal deformability $\Lambda$, which quantifies the induced quadrupole moment per unit external tidal field and is a key finite-size observable of compact stars in inspiralling binaries. We compute $\Lambda(M)$ for the same four EOS sets as in the preceding figures, using the formalism summarized in Appendix~\ref{sec:appendix_tidal}. Two robust trends emerge. First, for any given EOS, $\Lambda$ decreases monotonically with $M$ as stars become more compact. Second, at fixed $M$, a larger excluded-volume strength (larger $\kappa$) yields larger $\Lambda$, reflecting the stiffening of the EOS and the associated increase in stellar radius. Quantitatively, one may write $\Lambda=\tfrac{2}{3}\,k_2\,\mathcal{C}^{-5}\simeq \tfrac{2}{3}\,k_2\,(R/M)^5$, so that even modest radius variations at fixed mass can translate into sizeable changes in $\Lambda$ through its steep compactness dependence.}

The two classes of models exhibit distinct behaviors. For pure strange-quark stars [Fig.~\ref{fig:7}(a)], $\Lambda$ remains comparatively small across the mass range, reflecting their higher compactness. By contrast, self-bound hybrid stars [Figs.~\ref{fig:7}(b)–\ref{fig:7}(d)] typically produce larger $\Lambda$ at a fixed $M$, with the detailed level set by the transition density and by the repulsive excluded-volume correction: increasing $\kappa$ systematically shifts the curves upward, whereas varying the transition density alters the location and extent of the hybrid branch. 
For configurations featuring a sharp first--order interface and/or a finite surface density (self-bound stars), the tidal perturbation variable $y(r)$ is matched using the junction prescriptions detailed in Appendix~\ref{sec:appendix_tidal}, Eqs.~\eqref{eq:y_jump} and \eqref{eq:y_jump_at_surface}, before evaluating $k_2$ and $\Lambda$.

It is instructive to compare our $\Lambda(M)$ curves with the
constraint reported in Ref.~\cite{LIGOScientific:2017vwq}.  In the
low-spin scenario and assuming a uniform prior on the effective tidal
parameter $\tilde{\Lambda}$, that analysis yields a $90\%$ upper
limit $\tilde{\Lambda}\leq 800$; expanding $\Lambda(m)$ linearly
about $1.4\,M_\odot$ further gives
$\Lambda(1.4\,M_\odot)\leq 800$.  We adopt this single-star
threshold as an approximate observational reference and indicate it
by the gray markers in each panel of Fig.~\ref{fig:7}.
For the pure strange-quark-star sequences
[Fig.~\ref{fig:7}(a)], the models with $\kappa=0$ and
$\kappa=0.3$ predict $\Lambda(1.4\,M_\odot)$ comfortably
below the observational ceiling, reflecting the high compactness
characteristic of self-bound $uds$ stars.  The stiffest case,
$\kappa=0.6$, yields a somewhat larger $\Lambda$ but still
remains compatible with the bound.
In the self-bound hybrid case, the constraint is considerably more
restrictive.  In panels~\ref{fig:7}(b) and \ref{fig:7}(c), only the
softest parametrization ($\kappa=0$) satisfies
$\Lambda(1.4\,M_\odot)\lesssim 800$, and it does so only marginally,
with the curve lying just below the observational threshold.  Both
$\kappa=0.3$ and $\kappa=0.6$ exceed the bound in
these panels, the latter by a wide margin.  The situation improves
markedly for the high-transition-density set
[panel~\ref{fig:7}(d)], where the later onset of the $uds$ core
keeps the stellar radius smaller at $1.4\,M_\odot$: here both
$\kappa=0$ and $\kappa=0.3$ lie well below $800$, while only the
$\kappa=0.6$ models violate the constraint.
Physically, these trends reflect the steep compactness dependence
of the tidal deformability,
$\Lambda\propto\mathcal{C}^{-5}$: a stiffer EOS inflates the
stellar radius, and even a modest increase is amplified into a
large enhancement of $\Lambda$.  Notably, in
panels~\ref{fig:7}(b)–\ref{fig:7}(c) the $\kappa=0$ models are the
only ones compatible with the GW170817 tidal bound, yet these same
parametrizations were already ruled out in the mass–radius analysis
of Fig.~\ref{fig:6} for failing to support $2\,M_\odot$ stars.
This combined tension disfavors the low- and
intermediate-transition-density EOS sets for the self-bound hybrid
scenario.  In contrast, the high-transition-density set
[panel~\ref{fig:7}(d)] with $\kappa=0.3$ simultaneously satisfies
the tidal constraint and the $2\,M_\odot$ requirement, confirming
the conclusions drawn from Fig.~\ref{fig:6} and identifying this
region of parameter space as the most observationally viable for
self-bound hybrid configurations.

\begin{figure}[tb]
\centering
\includegraphics[width=0.99\linewidth]{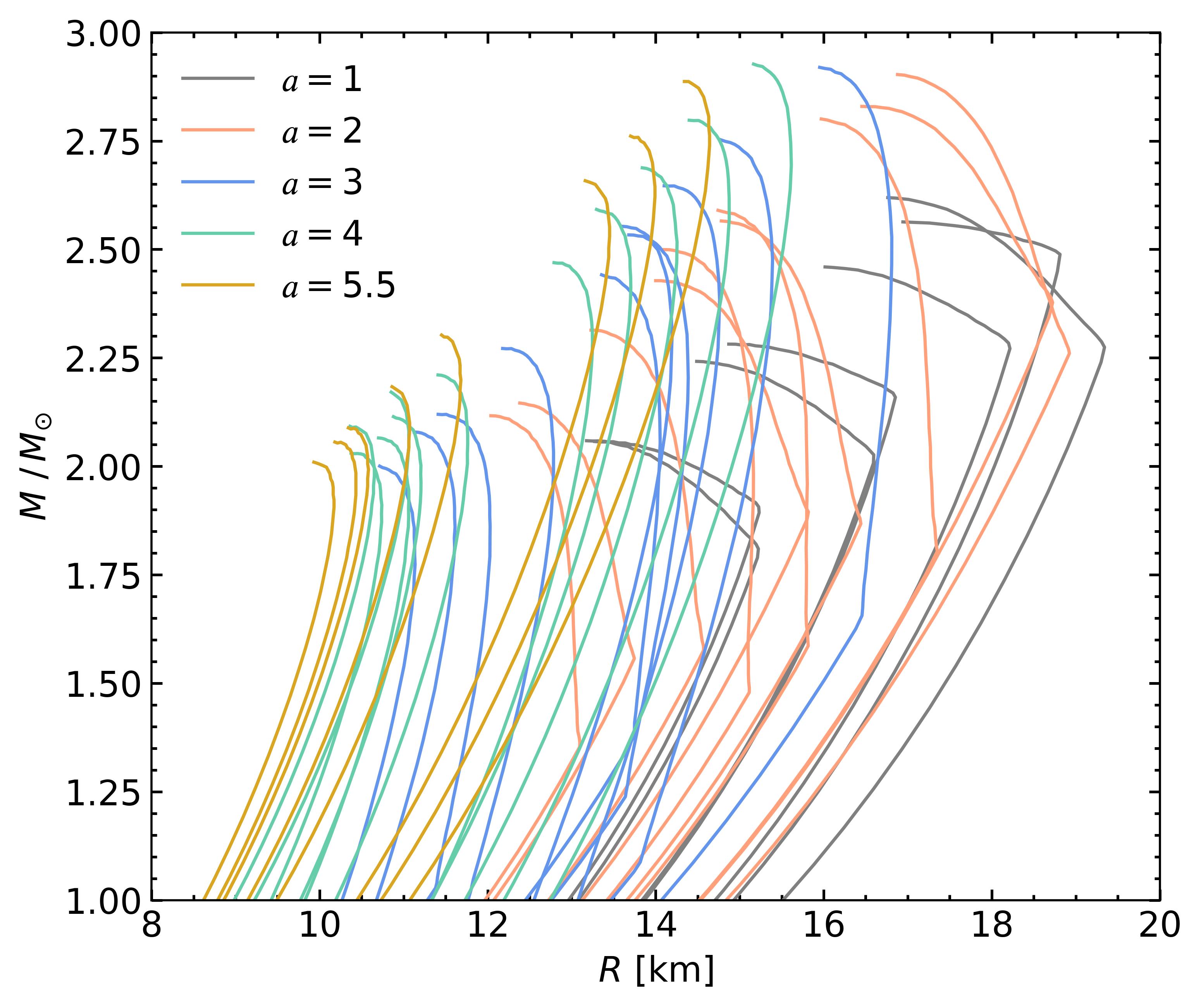}
\caption{Gravitational mass versus stellar radius for different choices of the EOS parameters ($a, C, \kappa$). For each $a$, the parameter $C$ was varied within its stability window (cf. Fig.~\ref{fig:1}) and the excluded-volume parameter was set to $\kappa \in \{0, 0.3, 0.6\}$.}
\label{fig:9}
\end{figure}

We also compute the stellar moment of inertia $I$ within the slow-rotation (Hartle–Thorne) approximation; the formalism is summarized in Appendix~\ref{sec:appendix_inertia}. Figure~\ref{fig:8} shows $I$ as a function of gravitational mass $M$ for the models of the preceding figures, together with two hadronic benchmarks (SLy and APR).

The effect of the excluded-volume parameter is clear. At fixed $M$, larger $\kappa$ yields larger $I$, reflecting EOS stiffening and the accompanying increase in radius. To leading order, $I$ scales approximately as $I\!\sim\! M R^2$; thus the radius growth driven by repulsion translates directly into a higher moment of inertia.
In Fig.~\ref{fig:8}(a), the pure SQM sequences track the hadronic fit rather closely, despite the different microphysics.
For self-bound hybrid stars [Figs.~\ref{fig:8}(b)–\ref{fig:8}(d)], the appearance of a three-flavor ($uds$) core leaves a distinct imprint on $I(M)$ at $p_c=p_\mathrm{tr}$. The behavior beyond the transition depends on the transition density. In the high–transition-density case [Fig.~\ref{fig:8}(b)], a “backbending’’ segment occurs: when a $uds$ stellar core forms, $I$ decreases even as $M$ continues to increase, indicating pronounced central condensation. By contrast, for intermediate and low transition densities [Figs.~\ref{fig:8}(c)–\ref{fig:8}(d)] the hybrid branch remains monotonic, with $I$ increasing with $M$ up to nearly the maximum-mass configuration (very close to $M_{\max}$ a change of trend can occur). Only dynamically stable configurations are shown.
Relative to SLy and APR, the $\kappa=0$ quark matter sequences lie closest to the hadronic curves, whereas strong repulsion ($\kappa=0.6$) produces substantially larger $I$ at a given $M$. These qualitative differences suggest that precision measurements of the moment of inertia (e.g., in double-pulsar systems) could help discriminate quark compositions and constrain the strength of repulsive interactions.

\section{Universal relations}
\label{sec:universal_relations}

One of the most active directions in compact-star physics is the search for \emph{universal relations}—functional correlations among macroscopic stellar quantities that depend only weakly on the underlying EOS. Such relations enable robust parameter inference: once some observables are measured, others can be estimated without detailed knowledge of the microphysics in the stellar core. Figure~\ref{fig:9} displays the mass–radius sequences for a representative set of EOS constructed by fixing five values of the confinement parameter \(a\) and varying \(C\), while the excluded-volume parameter is restricted to the three values \(\kappa = 0.0, 0.3, 0.6\). We retain only those models whose maximum mass lies between \(2\,M_\odot\) and \(3\,M_\odot\), and whose maximum radius remains below \(20\ \mathrm{km}\). The lower bound enforces consistency with \(\sim 2\,M_\odot\) pulsars, whereas the upper bound together with the \(20\ \mathrm{km}\) radius limit serve as conservative cuts to discard models with unrealistically large maximum masses and radii. This selected ensemble of models will be employed in the subsequent analysis of universal relations.

\begin{figure}[tb]
    \centering
    \includegraphics[width=0.99\linewidth]{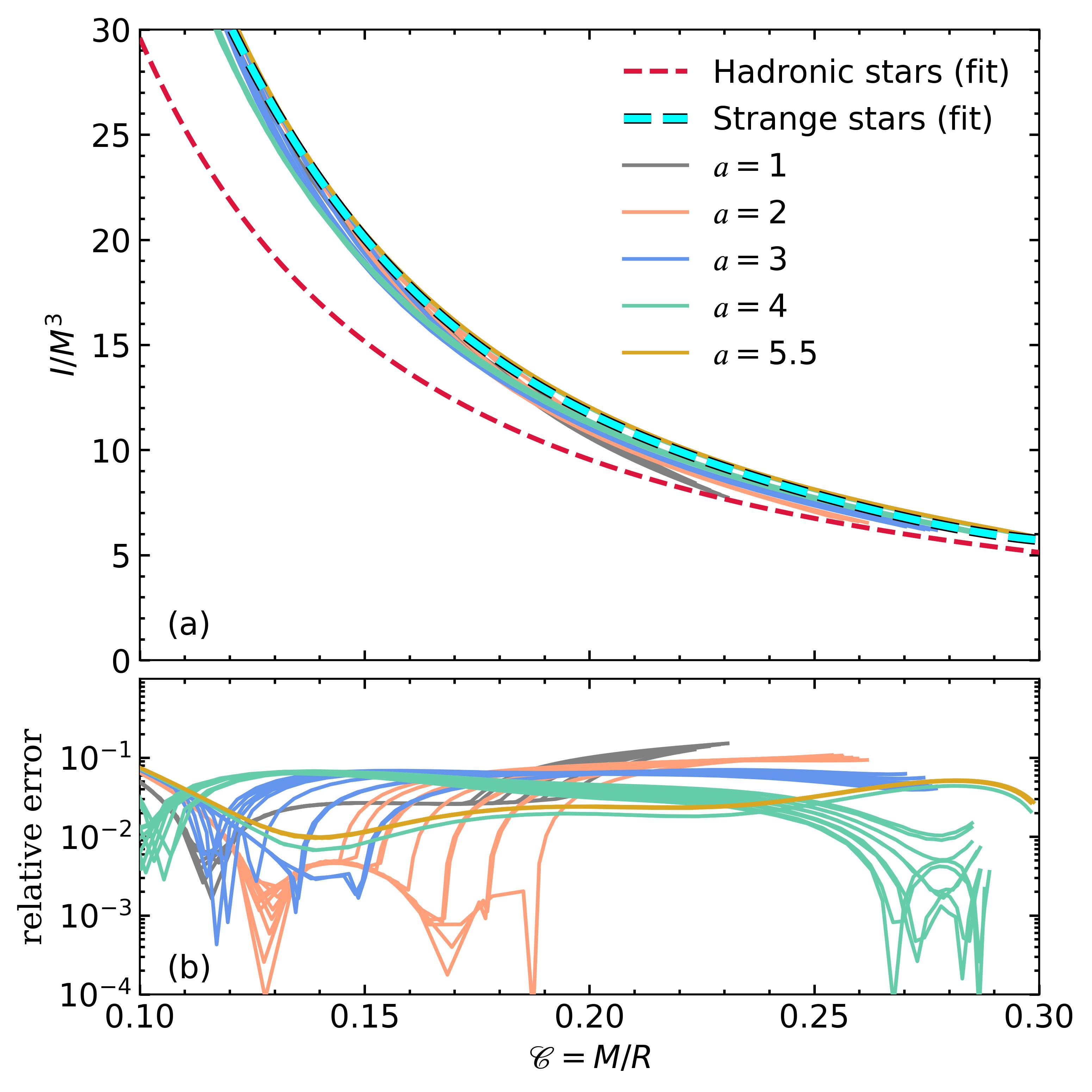}
\caption{(a) Dimensionless moment of inertia \(I/M_G^{3}\) vs. compactness \(M_G/R\) for sequences with different values of \(a\). Only EOSs with a maximum mass \(2 \le M_{\max}/M_\odot \le 3\) are included. The dashed lines are the global fits for strange quark stars~\cite{Lugones2025universality} and hadronic stars~\cite{Breu:2016ufb}. (b) Relative error, $\bigl|1-(I/M_G^{3})_{\rm num}/(I/M_G^{3})_{\rm fit}\bigr|$, with respect to the strange quark star fit.}    
\label{fig:10}
\end{figure}

\subsection{Dimensionless moment of inertia}

Figure~\ref{fig:10} shows the dimensionless moment of inertia, $\bar I \equiv I/M^{3}$, as a function of the compactness $\mathcal{C}\equiv M/R$.
Despite the strong $\kappa$-dependence seen in individual $M\!-\!R$ and $M\!-\!I$ relations, its imprint is nearly washed out in the $\bar I$–$\mathcal{C}$ plane: for fixed $(a,C)$, the three $\kappa$ tracks collapse to an excellent approximation onto a single curve. Moreover, the global trend across distinct $(a,C)$ families is captured by a single fit (cyan dashed line). Remarkably, the same fitting form successfully described strange stars in previous work~\cite{Lugones2025universality} based on the flavor-blind QMDD and vector MIT bag models. Its effectiveness here, for stars with different EOS ingredients and internal compositions, underscores its high degree of universality.

A noteworthy exception occurs over an intermediate range of compactness, where sequences with small $a$ (notably $a=1$ and $a=2$) depart more visibly from the fit, yielding the larger residuals shown in the lower panel. This behavior reflects the fact that for small $a$ values, the onset of the hybrid branch causes a significant reduction in stellar radii. This, in turn, produces a sharp drop in compactness (cf. Fig.~\ref{fig:6}) without a comparable change in the dimensionless moment of inertia. Outside that interval, the scatter about the fit remains small and approximately uniform across the different $(a,C)$ families.

It is worth emphasizing that the $\bar I$–$\mathcal{C}$ relation for self-bound quark stars found here differs markedly from the purely hadronic relation of Breu and Rezzolla~\cite{Breu:2016ufb}. Consequently, a simultaneous measurement of $M$, $R$, and $I$ would place a source at distinct loci in the $\bar I$–$\mathcal{C}$ plane, offering a clean diagnostic to discriminate a self-bound quark star from a conventional hadronic star.

\subsection{Gravitational versus baryonic compactness}

\begin{figure}[tb]
    \centering
    \includegraphics[width=0.99\linewidth]{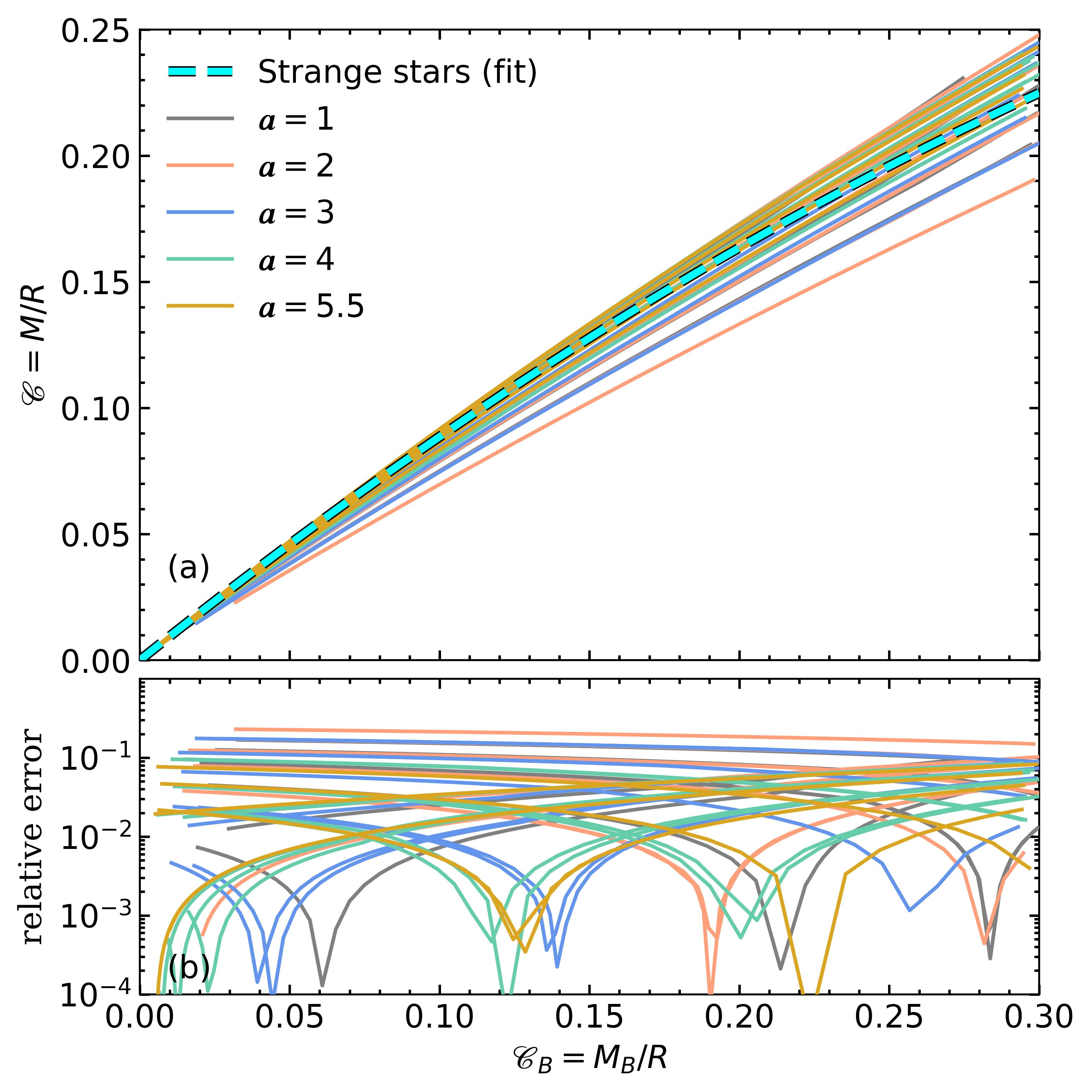}
\caption{Gravitational compactness \(\mathcal{C}\) versus baryonic compactness \(\mathcal{C}_B\) for stellar sequences labeled by the parameter \(a\) (colors as in the legend). 
Panel (a): solid curves show the models, while the cyan dashed line is the strange–quark–star fit \cite{Lugones2025universality}.
Panel (b): relative error with respect to the fit, \(|1-\mathcal{C}/\mathcal{C}^{\mathrm{fit}}|\).}
\label{fig:12}
\end{figure}

Figure~\ref{fig:12}(a) exhibits an approximately universal correlation for self-bound stars between the gravitational compactness $\mathcal{C} \equiv M/R$ and the baryonic compactness $\mathcal{C}_B \equiv M_B/R$, where the latter is computed from the conserved baryon number as given in Eqs.~(\ref{eq:AB_integral}) and (\ref{eq:baryon_mass_B7}). 
For a fixed $(a,C)$ family the three $\kappa$ sequences collapse onto essentially the same path, mirroring our findings for the $\bar{I}$--$\mathcal{C}$ and $\Lambda$--$\mathcal{C}$ relations. Tracks from distinct $(a,C)$ families cluster around a single global law, albeit with a larger dispersion than in the other universal relations, indicating an approximately EOS-insensitive mapping between $\mathcal{C}_B$ and $\mathcal{C}$.

To quantify this behavior, Fig.~\ref{fig:12}(a) includes the analytic fit introduced in Ref.~\cite{Lugones2025universality}, which accurately describes the combined vMIT and flavor-blind QMDD strange-star dataset. The accuracy is assessed in Fig.~\ref{fig:12}(b), which presents the relative deviation from that fit. For many parameterizations the scatter remains below the percent level, whereas for others it can reach up to $\sim 20\%$, mirroring the residual patterns seen in the $\bar{I}$--$\mathcal{C}$ and $\Lambda$--$\mathcal{C}$ cases.

In summary, Fig.~\ref{fig:12} shows that: (i) self-bound stars follow a $\mathcal{C}$--$\mathcal{C}_B$ correlation broadly consistent with our previously proposed fit; (ii) residuals are typically below the few-percent level across much of the explored range; and (iii) the $\mathcal{C}\!\leftrightarrow\!\mathcal{C}_B$ mapping provides a practical leading-order approximation to estimate $M_B$ and binding energies from $(M,R)$ (or $\mathcal{C}$).

\section{Conclusions}
\label{sec:conclusions}

We have investigated the macroscopic properties of self–bound quark stars within a flavor–dependent QMDD framework with an excluded–volume correction, exploring both SQM and self–bound two–flavor matter that undergoes a first–order transition to three flavors at high pressure. 
First, we mapped the $(a,C)$ parameter space at zero pressure to identify the parametrizations that produce self–bound matter, including those that undergo a first–order $ud\!\to\!uds$ transition. We then computed cold, $\beta$–equilibrated, charge–neutral stellar sequences by solving the TOV equations. The principal outcomes are as follows.

\begin{itemize}

\item[(i)] \textbf{Phase structure and stability.} {Over broad regions of the $(a,C)$ plane, the flavor--dependent QMDD EOS develops a first--order $ud\!\to\!uds$ transition. For the benchmark families analyzed in detail, the transition pressure $p_{\rm tr}$ shifts to lower values as the excluded--volume parameter $\kappa$ is increased. The corresponding energy--density discontinuities remain below Seidov’s onset--instability threshold: the Seidov diagnostic
$\gamma\equiv(\epsilon_{+}/\epsilon_{-})\big/\bigl[\tfrac{3}{2}\bigl(1+p_{\rm tr}/\epsilon_{-}\bigr)\bigr]$
lies in the range $0.85$--$0.97$ for all cases reported in Table~\ref{table_2} (cf.\ Eq.~\eqref{eq:seidov_parameter}), and is essentially insensitive to $\kappa$ at fixed $(a,C)$. Consistently, the full general--relativistic equilibrium sequences exhibit stable hybrid self--bound branches (identified by the standard turning--point criterion along each one--parameter family) in the mass--radius relation. }

\item[(ii)] \textbf{Global stellar properties and observational constraints.} 
{Over a broad region of the EOS parameter space, the two-to-three flavor transition sets in \emph{before} the maximum-mass configuration is reached, i.e., $p_c^{\max}\!\ge\!p_{\rm tr}$ (equivalently, the $uds$ core forms along the stable branch prior to $M_{\max}$). As a result, self-bound hybrid stars constitute a generic outcome of the flavor-dependent scheme, and the onset of the $uds$ core produces a characteristic \emph{kink} in the stellar sequences at $p_c=p_{\rm tr}$, visible in the $M$--$R$, $\Lambda(M)$, and $I(M)$ curves. }
The parameter $\kappa$ has a strong and systematic impact on the stiffness of the EOS and therefore on masses, radii, tidal deformabilities, and moments of inertia. In the absence of excluded volume ($\kappa=0$), maximum masses tend to undershoot the $\sim 2\,M_\odot$ constraint, whereas large $\kappa$ values can overshoot radius bounds at canonical masses.
The tidal deformability provides an additional and complementary filter.
Comparing our $\Lambda(M)$ curves with the GW170817 low-spin upper
limit $\Lambda(1.4\,M_\odot)\leq 800$~\cite{LIGOScientific:2017vwq},
we find that pure strange-quark-star sequences satisfy the bound for
all explored values of $\kappa$, owing to their intrinsically high
compactness.  For self-bound hybrid stars the constraint is more
selective: at low and intermediate transition densities, only the
$\kappa=0$ models comply with the tidal bound, and they do so only
marginally—yet these same parametrizations fail to reach
$2\,M_\odot$, effectively ruling out those EOS sets.  The tension is
resolved for high-transition-density configurations, where
$\kappa=0.3$ simultaneously satisfies $\Lambda(1.4\,M_\odot)\lesssim
800$ and $M_{\max}\gtrsim 2\,M_\odot$, while also remaining
consistent with NICER radius inferences~\cite{Miller2019, Riley2019,
Miller2021, Riley2021} and the compact object in
HESS~J1731$-$347~\cite{Doroshenko2022}.  Taken together, the
mass, radius, and tidal constraints converge on intermediate repulsion
($\kappa\approx 0.3$) with a high transition density as the most
observationally viable region of parameter space for self-bound hybrid
configurations.

\item[(iii)] \textbf{Universal relations.} Two EOS–insensitive trends emerge:
    (a) the dimensionless moment of inertia $\bar I\!\equiv\!I/M_G^3$ versus compactness $\mathcal{C}$, and
    (b) the gravitational versus baryonic compactness, $\mathcal{C}$–$\mathcal{C}_B$.
    In both cases, the explicit dependence on $\kappa$ largely disappears once the relations are expressed in dimensionless form. These relations follow the flavor–blind strange–quark–star fits reported previously \cite{Lugones2025universality}, with typical residuals below the percent level and larger deviations up to $\sim 10$–$20\%$ for a few $a$ families (where the phase transition reshapes the stellar structure). The $\mathcal{C}$–$\mathcal{C}_B$ mapping is less tight. Self-bound hybrid stars remain close to the universal relations but exhibit slightly larger departures than strange quark stars, due to the change in stiffness across the two- to three-flavor EOS transition. Because the onset of this transition varies among parameter sets, the effect manifests as a spread of curves for some parametrizations.

\item[(iv)] \textbf{Quark--hadron discrimination.} For fixed compactness, self--bound stars follow $\bar I$--$\mathcal{C}$ systematics that can differ from hadronic fits, reflecting their distinct surface boundary condition (finite surface density) and more homogeneous interiors. In practice, however, the separation is not uniform: in the high--compactness regime most relevant for heavy stars the differences with respect to hadronic trends can become comparatively modest. Anyway, joint constraints on $(M,R)$ together with an independent moment--of--inertia measurement may help discriminate self--bound quark stars from purely hadronic stars, but this test will likely require sufficiently precise data. We also find that models with $\kappa=0$ yield $I(M)$ closer to representative hadronic benchmarks (e.g., SLy and APR), whereas increasing $\kappa$ systematically raises $I$; this trend could be leveraged by future moment--of--inertia measurements.

\end{itemize}

Overall, the flavor--dependent QMDD model with excluded volume accommodates both SQM stars and self--bound hybrid stars compatible with current astrophysical constraints, and it exhibits EOS--insensitive trends that can be useful for inference within the self--bound sector. In particular, the $\bar I$--$\mathcal{C}$ relation is comparatively tight across our parameter space, whereas the $\mathcal{C}$--$\mathcal{C}_B$ mapping---though practical for first--order estimates of baryonic masses and binding energies---shows a larger, model--dependent scatter in specific regions of parameter space.

Taken together, these results provide a minimal and coherent baseline for interpreting self–bound configurations within a simple, transparent framework. The dimensionless relations can be folded into inference pipelines either as informative priors or as independent consistency checks, by comparing the quantities inferred from data with the values predicted by our fits.

\section*{Acknowledgements}

GL acknowledges the financial support from the Brazilian agency CNPq (grant 316844/2021-7). AGG acknowledges the financial support from CONICET under Grant No. PIP 22-24 11220210100150CO,  ANPCyT (Argentina) under Grant PICT20-01847, and the National University of La Plata (Argentina), Project No. X960.

\appendix

\section{The EOS}
\label{sec:appendix_EOS}

\subsection{EOS for pointlike quasiparticles}

At $T=0$, the energy density, pressure, and chemical potentials of \emph{pointlike} ($\mathrm{pl}$) quasiparticles are written in terms of the number densities $n_i$ as
\begin{eqnarray}
\epsilon_\mathrm{pl} &=&  \sum_{i} \epsilon_{\mathrm{pl},i} + \epsilon_e    =  \sum_{i} g M_i^4 \chi(x_i) + \epsilon_e, \label{eq:pressure_general} \\ 
p_\mathrm{pl} &=&   \sum_{i} p_{\mathrm{pl},i} + p_e  = \sum_{i}  \left(  g M_i^4 \phi\left(x_i\right)-B_i \right) + p_e, \qquad \label{eq:epsilon_general} \\
\mu_{\mathrm{pl},i} &=& M_i \sqrt{x_i^2+1}-\frac{ B_i }{ n_i} .   \label{eq:mu_general}
\end{eqnarray}
Here $g$ denotes the spin–color degeneracy (for quarks $g=6$), $M_i$ is the effective quasiparticle mass for flavor $i$, and the quantity $B_i$ is an \emph{effective bag term} arising from the density dependence of $M_i$:
\begin{equation}
-\,B_i \;=\; g\,\beta(x_i)\,M_i^3\,n_i\,\frac{\partial M_i}{\partial n_i}.
\label{eq:bag_term}
\end{equation}
The dimensionless Fermi momentum is
\begin{equation}
x_i \;=\; \frac{1}{M_i}\!\left(\frac{6\pi^2 n_i}{g}\right)^{\!1/3},
\end{equation}
and 
\begin{align}
\chi(x) &= \frac{x\sqrt{1+x^2}\,(2x^2+1) - \sinh^{-1}x}{16\pi^2}, \\
\phi(x) &= \frac{x\sqrt{1+x^2}\,(2x^2-3) + 3\sinh^{-1}x}{48\pi^2}, \\
\beta(x) &= \frac{x\sqrt{1+x^2} - \sinh^{-1}x}{4\pi^2}.
\end{align}
Electron contributions $(\epsilon_e,p_e)$ follow the usual zero-temperature expressions with $g=2$.

\subsection{Excluded–volume effects}
\label{sec:exclude_volume}

We incorporate short–range repulsion via an excluded–volume prescription in the $T=0$ Helmholtz representation, following Refs.~\cite{Lugones:2023zfd,Lugones:2024ryz}. Consistent with the flavor–dependent mass model, we take
\begin{equation}
b_i \;=\; \frac{\kappa}{n_i}\,, \qquad \kappa>0,
\end{equation}
so that each flavor removes a fraction $b_i n_i=\kappa$ per unit volume. Thus the \emph{available} volume fraction is flavor–independent,
\begin{equation}
q \;\equiv\; 1-\kappa\,, \qquad 0<q\le 1,
\end{equation}
even though the ansatz is flavor–dependent through $n_i$. In practice, the excluded–volume mapping amounts to evaluating the point–like quark expressions at rescaled densities $\tilde n_i\equiv n_i/q$ and multiplying the quark–sector thermodynamic quantities by $q$ (the electron sector is left unchanged). The excluded–volume corrected EOS reads
\begin{eqnarray}
\epsilon & = &  \sum_{i}  q \epsilon_{\mathrm{pl},i}    + \epsilon_e = \sum_{i} q g \tilde{M}_i^4 \chi\left(\tilde{x}_i\right)+\epsilon_e,   \label{eq:epsilon_with_q}\\
p & = &   \sum_{i}  q p_{\mathrm{pl},i}    +  p_e =  \sum_{i}  q  (g \tilde{M}_i^4 \phi\left(\tilde{x}_i\right)-\tilde{B}_i ) + p_e,  \quad \qquad \label{eq:p_with_q} \\ 
\mu_i & = & \tilde{M}_i \sqrt{\tilde{x}_i^2+1}-\frac{\tilde{B}_i}{ \tilde{n}_i}   ,
\end{eqnarray}
with
\begin{eqnarray}
\tilde{n}_i & = & n_i / q ,\\
\tilde{M}_i & = & M_i\left(\tilde{n}_i\right), \\
\tilde{x}_i & = & \frac{1}{\tilde{M}_i}\left[\frac{6 \pi^2\left(n_i / q\right)}{g}\right]^{1 / 3} , \\
-\tilde{B}_i & = & g \beta\left(\tilde{x}_i\right) \tilde{M}_i^3 \tilde{n}_i \frac{\partial \tilde{M}_i}{\partial \tilde{n}_i} .
\end{eqnarray}
Thermodynamic consistency is preserved by construction, and the point–like limit is recovered for
$\kappa\to 0$ ($q\to 1$). We note that the Gibbs free energy per baryon at $p=0$ is insensitive to
$\kappa$ in this construction, in agreement with the results shown in the main text.

\subsection{Compact-star conditions}

For cold, catalyzed matter in compact stars we impose local charge neutrality and $\beta$ equilibrium (with freely streaming neutrinos). The chemical-equilibrium conditions read
\begin{align}
\mu_d = \mu_u + \mu_e, \qquad
\mu_s &= \mu_d, 
\end{align}
with $\mu_{\nu_e}=0$, while charge neutrality requires
\begin{equation}
\tfrac{2}{3}\,n_u - \tfrac{1}{3}\,n_d - \tfrac{1}{3}\,n_s - n_e \;=\; 0.
\end{equation}
(Throughout, we use natural units $\hbar=c=1$.)

\section{Stellar structure}
\label{sec:appendix_stellar_structure}

The EOS developed above is used to determine global properties of self–bound compact stars—mass–radius relations, tidal deformabilities, and (in the slow–rotation regime) moments of inertia. For completeness, we summarize here the field equations and integrals needed for these calculations. Throughout this appendix we work in geometrized units, $G=c=1$.

\subsection{TOV equations}
\label{sec:appendix_TOV}

We assume a static, spherically symmetric spacetime with line element
\begin{equation}
ds^2 = e^{2\nu(r)} dt^2 - e^{2\lambda(r)} dr^2 - r^2 \bigl(d\theta^2 + \sin^2\theta\, d\phi^2\bigr),
\end{equation}
and a perfect–fluid stress–energy tensor characterized by energy density $\epsilon(r)$ and pressure $p(r)$. Hydrostatic equilibrium is governed by the TOV equations
\begin{align}
\frac{dp}{dr} &= -\,\frac{\bigl[\epsilon(r)+p(r)\bigr]\bigl[m(r)+4\pi r^3 p(r)\bigr]}
{r\bigl[r-2m(r)\bigr]}\,,
\label{eq:TOV_dp}\\[3pt]
\frac{dm}{dr} &= 4\pi r^2 \epsilon(r)\,,
\label{eq:TOV_dm}\\[3pt]
\frac{d\nu}{dr} &= \frac{m(r)+4\pi r^3 p(r)}{r\bigl[r-2m(r)\bigr]}\,,
\label{eq:TOV_dnu}
\end{align}
where $m(r)$ is the enclosed gravitational mass. Regularity at the center requires $m(0)=0$ and $p(0)=p_c$ (the latter sets the stellar model). The stellar surface $r=R$ is defined by $p(R)=0$, and the gravitational mass is
\begin{equation}
M \equiv m(R) .
\end{equation}
Matching to the exterior Schwarzschild solution fixes $e^{2\nu(R)} = 1-2M/R$.
The compactness used in the main text is $\mathcal{C}\equiv M/R$.

Let $n_B(r)$ denote the proper baryon number density. The total baryon number is
\begin{equation}
A_B = \int_0^R 4\pi r^2 n_B(r)\,\left(1-\frac{2m(r)}{r}\right)^{-1/2} dr\,,
\label{eq:AB_integral}
\end{equation}
and the baryonic mass is $M_B \equiv m_B A_B$, where $m_B$ is a chosen baryon rest mass
(often taken as the neutron mass $m_n$). For orientation, a canonical compact star with
$M\simeq 1.4\,M_\odot$ typically has $A_B\simeq 1.7\times 10^{57}$, implying
\begin{equation}
M_B= A_B m_n \simeq 0.842 A_{B57} M_\odot,
\quad A_{B57}\equiv \frac{A_B}{10^{57}}.
\label{eq:baryon_mass_B7}
\end{equation}
In the absence of mass accretion or mass loss, $A_B$ (and hence $M_B$) is conserved along
quasi–static evolutions.

\subsection{Dimensionless tidal deformability}
\label{sec:appendix_tidal}

During the inspiral of a binary neutron–star system, the external quadrupolar tidal field
$\mathcal{E}_{ij}$ induces a mass–quadrupole moment $Q_{ij}$ in each star. In linear
response one writes
\begin{equation}
Q_{ij} = -\lambda\mathcal{E}_{ij} = -\Lambda M^{5}\mathcal{E}_{ij},
\end{equation}
where $\lambda$ is the (dimensionful) tidal deformability, $M$ is the gravitational mass,
and $\Lambda \equiv \lambda/M^{5}$ is the \emph{dimensionless} tidal deformability quoted
in the main text. It is given by
\begin{equation}
\Lambda = \frac{2}{3}k_{2}\mathcal{C}^{-5},
\label{eq:definition_of_Lambda}
\end{equation}
being $k_{2}$ the (quadrupolar) Love number.

Following the standard relativistic perturbation formalism, $k_{2}$ can be expressed in
terms of $\mathcal{C}$ and the surface value $y_{R}\equiv y(r=R)$ of the first order
metric perturbation function $y(r)$:
\begin{equation}
\begin{aligned}
k_{2} =& \frac{8}{5}\mathcal{C}^{5}(1-2\mathcal{C})^{2}
\frac{ \bigl[2 - y_{R} + 2\mathcal{C}(y_{R}-1)\bigr] }
{ \vphantom{\Big|}\mathcal{D} },
\end{aligned}
\end{equation}
with
\begin{equation}
\begin{aligned}
\mathcal{D} =&
\; 2\mathcal{C}\Bigl[6 - 3y_{R} + 3\mathcal{C}(5y_{R}-8)\Bigr]
\\
&+ 4\mathcal{C}^{3}\Bigl[13 - 11y_{R}
    + \mathcal{C}(3y_{R}-2) + 2\mathcal{C}^{2}(1+y_{R})\Bigr]\\
&+ 3(1-2\mathcal{C})^{2}\Bigl[2 - y_{R} + 2\mathcal{C}(y_{R}-1)\Bigr]  \ln(1-2\mathcal{C}).
\end{aligned}
\end{equation}
The function $y(r)$ obeys the first–order ODE
\begin{equation}
r\frac{dy}{dr} + y^{2} + yF(r) + r^{2} Q(r) = 0,
\label{EL15}
\end{equation}
with coefficients
\begin{align}
F(r) &= \frac{1 - 4\pi r^{2}[\epsilon(r)-p(r)]}{1 - 2m(r)/r},
\\[3pt]
Q(r) &= \frac{4\pi\!\left[5\epsilon(r)+9p(r)
      + \frac{\epsilon(r)+p(r)}{c_{s}^{2}(r)} - \frac{6}{4\pi r^{2}} \right]}
      {1 - 2m(r)/r}
\nonumber\\
&\quad -\frac{4m^{2}(r)}{r^{4}}
      \frac{\bigl[1 + 4\pi r^{3}p(r)/m(r)\bigr]^{2}}
           {\bigl[1 - 2m(r)/r\bigr]^{2}},
\label{EL17}
\end{align}
where $c_{s}^{2}\equiv dp/d\epsilon$, and $m(r)$, $p(r)$, $\epsilon(r)$ are the TOV
solutions from Eqs.~\eqref{eq:TOV_dp}–\eqref{eq:TOV_dnu}. Regularity at the center sets
$y(0)=2$. The Love number $k_{2}$ and hence $\Lambda$ follow by integrating
\eqref{EL15} together with the TOV equations up to $r=R$. For additional details, see
Ref.~\cite{Postnikov:2010yn}.

{
\paragraph*{Junction conditions at a sharp interface.}
When a sharp ($ud\!\to\!uds$) first--order transition occurs at pressure $p_{\rm tr}$,
the interface is located at the radius $r_{\rm tr}$ defined by $p(r_{\rm tr})=p_{\rm tr}$.
Assuming an infinitesimally thin interface free of surface stresses, the Israel--Darmois
junction conditions imply continuity of the induced metric and extrinsic curvature.
In static spherical symmetry this translates into continuity of the metric functions
(hence $m$ and $\nu$) and of the pressure $p$, while allowing a jump in the energy density.
}

{
To avoid any sign ambiguity, we define the density jump as
\begin{equation}
\epsilon_{\rm in}\equiv \epsilon(r_{\rm tr}-0),\quad
\epsilon_{\rm out}\equiv \epsilon(r_{\rm tr}+0),\quad
\Delta\epsilon \equiv \epsilon_{\rm in}-\epsilon_{\rm out} > 0,
\end{equation}
where ``in'' (``out'') denotes the inner (outer) side of the interface, i.e.\ smaller (larger) $r$.
For the tidal sector, the first--order ordinary differential equation for $y(r)=rH'(r)/H(r)$ must be matched by
integrating the perturbation equation across the density discontinuity
\cite{Postnikov:2010yn, Takatsy:2020bnx}, which yields
\begin{equation}
\label{eq:y_jump}
y(r_{\rm tr}+0)=y(r_{\rm tr}-0)-
\frac{4\pi\,r_{\rm tr}^{3}\,\Delta\epsilon}{\,m(r_{\rm tr}) + 4\pi r_{\rm tr}^{3} p_{\rm tr}\,}\,.
\end{equation}
After enforcing \eqref{eq:y_jump}, the integration proceeds to the stellar surface in order
to compute $k_{2}$ and $\Lambda$. This prescription ensures continuity of the physical
(gauge--invariant) metric perturbations and a well--posed Love number.
}

{
\paragraph*{Surface correction for self--bound stars.}
For self--bound configurations with a finite surface density $\epsilon_s\equiv \epsilon(R^-)>0$,
the pressure vanishes at the surface, $p(R)=0$, while the exterior is vacuum, $\epsilon(R^+)=0$.
Equation~\eqref{eq:y_jump} then applies with $r_{\rm tr}=R$, $p_{\rm tr}=0$, $m(R)=M$, and
$\Delta\epsilon=\epsilon_s$, giving
\begin{equation}
y(R^+)=y(R^-)-\frac{4\pi R^3\,\epsilon_s}{M}\,.
\label{eq:y_jump_at_surface}
\end{equation}
The value entering the standard $k_2(\beta,y_R)$ formula is the exterior one, $y_R\equiv y(R^+)$.
}

\bigskip

\subsection{Moment of inertia for slow rigid rotation}
\label{sec:appendix_inertia}

We evaluate the stellar moment of inertia within the slow–rotation (Hartle–Thorne) framework, restricting attention to uniform rotation with angular frequency $\Omega$ as measured at infinity~\cite{Hartle:1967he}. To first order in $\Omega$, the total angular momentum is $J=\Omega\,I$, so that
\begin{equation}
I  =  \frac{J}{\Omega}\,,
\end{equation}
and $I$ is determined entirely by the mass–energy distribution and the spacetime of the corresponding nonrotating configuration.

It is convenient to compute $I$ by integrating a first–order equation for the cumulative (enclosed) moment of inertia $\mathcal{I}(r)$,
\begin{equation}
\frac{d\mathcal{I}}{dr}
= \frac{8\pi}{3}\, r^{4}\,\epsilon\!\left(1+\frac{p}{\epsilon}\right)
\!\left(1-\frac{5}{2}\frac{\mathcal{I}}{r^{3}}+\frac{\mathcal{I}^{2}}{r^{6}}\right)
\!\left(1-\frac{2m}{r}\right)^{-1},
\label{eq:moment_of_inertia}
\end{equation}
where $m(r)$ is the mass function, and $p(r)$ and $\epsilon(r)$ are supplied by the TOV solution (Sec.~\ref{sec:appendix_TOV}). Regularity at the center imposes $\mathcal{I}(0)=0$, and the stellar moment of inertia follows as
\begin{equation}
I  =  \mathcal{I}(R),
\end{equation}
with $R$ the radius where $p(R)=0$. This formulation is equivalent to the standard computation based on the frame–dragging function $\bar\omega(r)=\Omega-\omega(r)$ and its exterior matching; see, e.g., Refs.~\cite{Hu:2023vsq,Dong:2023vxv}.

\paragraph*{Remark on first–order phase transitions.}
Across an infinitesimal interface at $r_{\rm tr}$, one integrates Eq.~\eqref{eq:moment_of_inertia} on each side and impose continuity of the angular–momentum flux. In the equivalent frame–dragging formulation this means $\bar\omega$ and $r^{4}j\,\bar\omega'$ (with $j\!\equiv\!e^{-(\nu+\lambda)/2}$) are continuous at $r_{\rm tr}$. Hence $\mathcal{I}(r)$ is continuous and $I=\mathcal{I}(R)$ is well defined.

\bibliography{references}
\end{document}